\documentclass[useAMS,usenatbib]{mn2e}
\usepackage{txfonts}
\usepackage{graphicx}
\usepackage{verbatim}
\usepackage{rotating}
\usepackage[flushleft]{threeparttable}
\usepackage{color}
\usepackage{calc}% http://ctan.org/pkg/calc

\newcommand\finetilde{{\raise.17ex\hbox{$\scriptstyle\sim$}}} % a more appealing tilde for use in text

\begin{document}

\title[Scattering in Pulsar Timing Analysis]{Robust Estimation of Scattering in Pulsar Timing Analysis}
\author[L. Lentati et al.]{\parbox{\textwidth}{L. Lentati$^{1}$\thanks{E-mail:
ltl21@cam.ac.uk}, M. Kerr$^{2}$, S. Dai$^{2}$, R. M. Shannon$^{2,3}$, G. Hobbs$^{2}$, S. Os{\l}owski$^{4,5,6}$ }\vspace{0.4cm}\\ %
$^{1}$ Astrophysics Group, Cavendish Laboratory, JJ Thomson Avenue,  Cambridge, CB3 0HE, UK\\
$^{2}$ Australia Telescope National Facility, CSIRO Astronomy \& Space Science, P.O. Box 76, Epping, NSW 1710, Australia\\
$^{3}$ International Centre for Radio Astronomy Research, Curtin University, Bentley, Western Australia 6102, Australia\\
$^{4}$ Centre for Astrophysics and Supercomputing, Swinburne University of Technology, Post Office Box 218 Hawthorn, VIC 3122, Australia\\
$^{5}$ Fakult\"{a}t f\"{u}r Physik, Universit\"{a}t Bielefeld, Postfach 100131, 33501 Bielefeld, Germany\\
$^{6}$ Max Planck Institute for Radio Astronomy, Auf dem H\"{u}gel 69, D-53121 Bonn, Germany}

\maketitle

\label{firstpage}

\begin{abstract}%
We present a robust approach to incorporating models for the time-variable broadening of the pulse profile due to scattering in the ionized interstellar medium into profile-domain pulsar timing analysis.   We use this approach to simultaneously estimate temporal variations in both the dispersion measure (DM) and scattering, together with a model for the pulse profile that includes smooth evolution as a function of frequency, and the pulsar's timing model.  We show that fixing the scattering timescales when forming time-of-arrival estimates, as has been suggested in the context of traditional pulsar timing analysis,  can significantly underestimate the uncertainties in both DM, and the arrival time of the pulse, leading to bias in the timing parameters.  We apply our method using a new, publicly available, GPU accelerated code, both to simulations, and observations of the millisecond pulsar PSR J1643$-$1224.  This pulsar is known to exhibit significant scattering variability compared to typical millisecond pulsars, and we find including low-frequency ($< 1$\,GHz) data without a model for these scattering variations leads to significant periodic structure in the DM,  and also biases the astrometric parameters at the $4\sigma$ level,  for example, changing proper motion in right ascension by $0.50 \pm 0.12$.  If low frequency observations are to be included when significant scattering variations are present, we conclude it is necessary to not just model those variations, but also to sample the parameters that describe the variations simultaneously with all other parameters in the model, a task for which profile domain pulsar timing is ideally suited.
\end{abstract}

\maketitle


\section{Introduction}

The eventual detection of gravitational waves in the nanoHertz window using a pulsar timing array \citep{1990ApJ...361..300F} will require a thorough understanding of the myriad mechanisms that can impact either the shape, or the time of arrival (ToA) of the pulses of light from those pulsars.  The Ionised Interstellar Medium (IISM) is known to be the dominant such mechanism in pulsar timing experiments (e.g., \citealt{2016ApJ...819..155L}), and introduces \textit{both} changes in the shape of the profile, and delays in the arrival times.  

The delays in the arrival times are primarily the result of interstellar dispersion.  As the pulse propagates through the ionised plasma that makes up the IISM, it interacts with free electrons causing a frequency dependent delay.  This delay is proportional the the integrated column density of electrons along our line of sight to the pulsar, called the dispersion measure (DM), and scales as $\nu^{-2}$, with $\nu$ the observing frequency.  As our line of sight to the pulsar changes with time, so too does the observed column density, introducing a time-variable delay in arrival times.

Over the last several years significant progress has been made in modeling these variations in DM, and in propagating our uncertainties in that model through to the timing parameters, with multiple different methods in use by groups around the world (e.g. \citealt{2013MNRAS.429.2161K, 2013ApJ...762...94D, 2014MNRAS.437.3004L}).

A more subtle effect due to the IISM is that of scattering,  where inhomogeneities in the IISM cause both intensity variations (called diffractive scintillation, e.g. \citealt{1992RSPTA.341..151N}) and also a broadening of the pulse profile.  The impact of scattering, as for DM variations,  will also change with time as our line of sight to the pulsar changes, and in this paper we will be concerned with how to robustly incorporate scattering variations into pulsar timing analysis.

Given the complexity of the IISM, and the steep frequency dependence of the impact it has on pulsar timing, one can rightly ask why we should not just migrate to high-frequency observations alone.  Indeed, this approach has already resulted in the most sensitive limit on an isotropic gravitational wave background using a pulsar timing array to date \citep{2015Sci...349.1522S}.  The simple answer is that pulsars are known to have steep spectral indices ($S_{\nu} \propto \nu^{-1.8}$ on average, \citealt{2000A&AS..147..195M}),  and so are much brighter at lower observing frequencies.  Given the strong dependence of the detection probability of gravitational waves using a pulsar timing array on the number of pulsars in the array (e.g., \citealt{2016ApJ...819L...6T}), being able to include more pulsars by virtue of observing at lower frequencies where they are brightest has clear benefits.

Previous attempts to model scattering have either made significant assumptions about \textit{a priori} unknown quantities, such as the intrinsic pulse shape, or only consider scattering independent of other aspects of the model such as the intrinsic variability in the pulse shape, or DM variations.  In both cases the result is a failure to propagate the uncertainties from these other parameters into the scattering measurements.

\begin{figure*}
\begin{center}$
\begin{array}{cc}
\hspace{-0.5cm}
\includegraphics[trim = 65 10 30 0, clip,width=90mm]{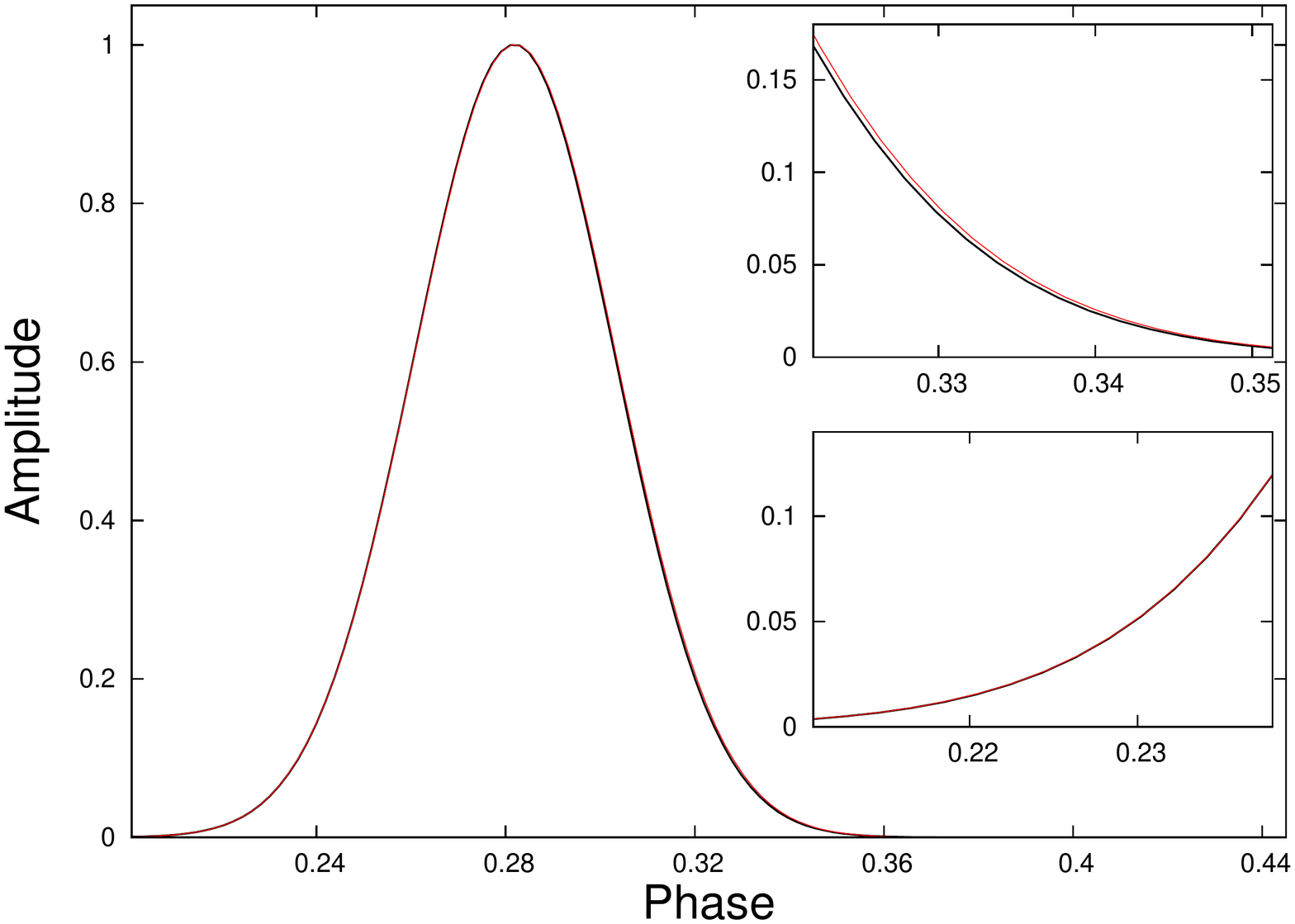} &
\hspace{-1.0cm}
\includegraphics[trim = 65 10 30 0, clip,width=90mm]{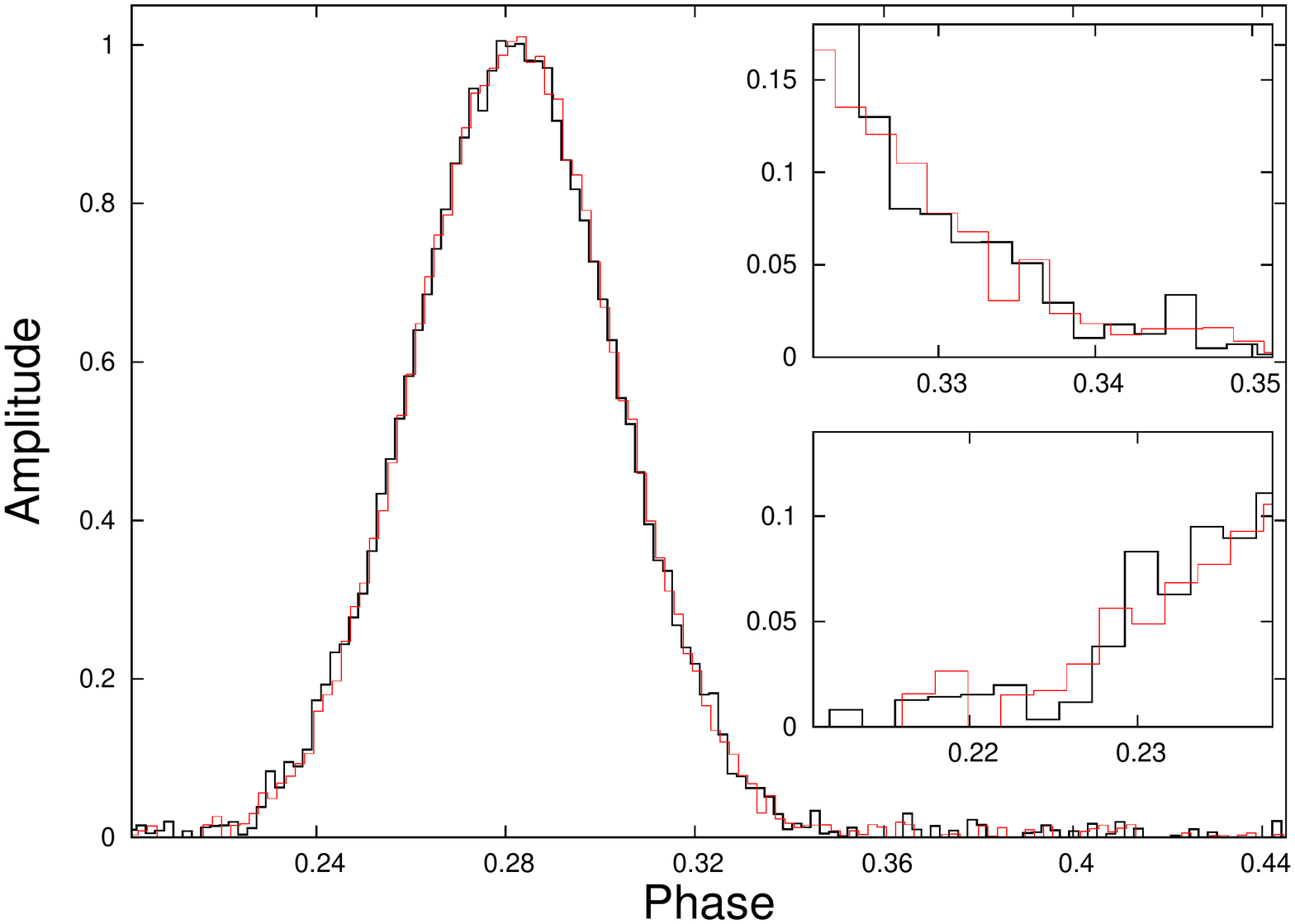}\\
\includegraphics[trim = 40 10 50 0, clip,width=100mm]{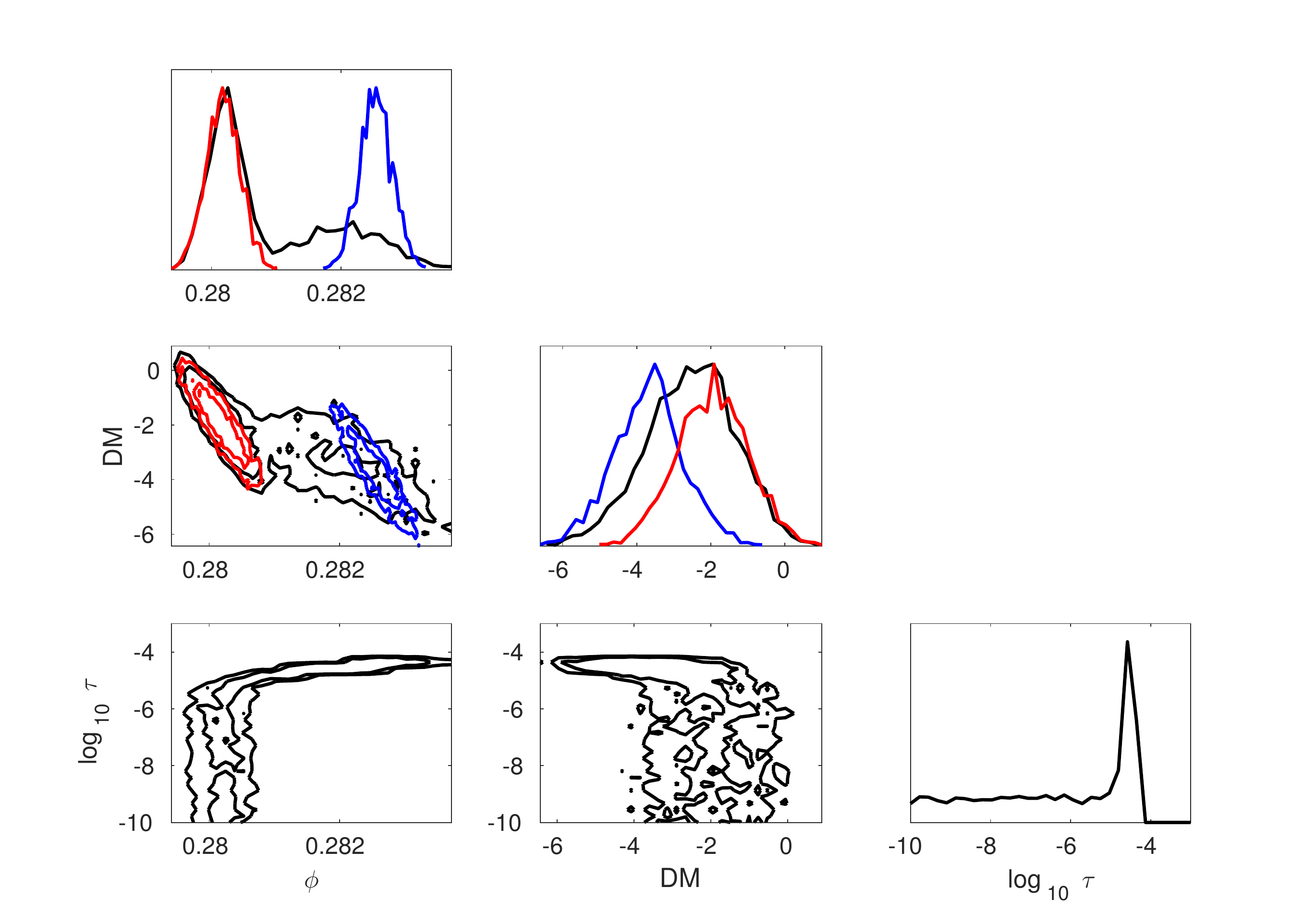} &
\includegraphics[trim = 35 0 25 0, clip,width=70mm]{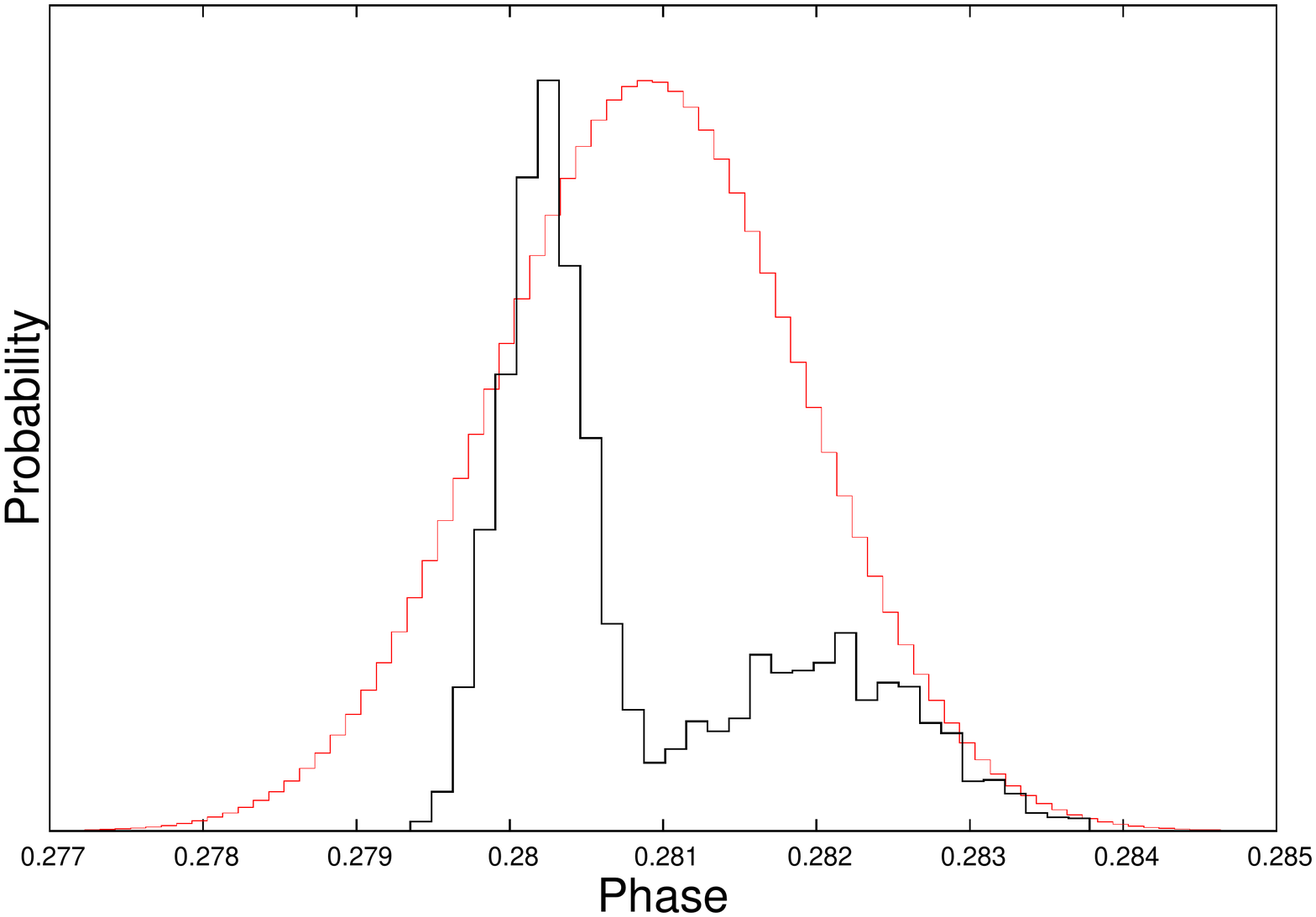}\\
\end{array}$
\end{center}
\caption{(Top panels) Simulated data before adding noise (left-panel) and after (right panel) for a Gaussian pulse profile with a scattering timescale of     $10^{-4.5}$\,s at a reference frequency of 1~GHz.  Sub-panels show zoomed in regions of the trailing- (top) and leading-edge (bottom).    (Bottom-left panel) One- and two-dimensional marginalised posterior distributions from a three-dimensional analysis of the simulated observation  (black lines).  We fit simultaneously for the ToA, the DM, and $\log_{10}$  of the scattering timescale.  We compare these parameter estimates with those obtained from a two-dimensional analysis where we fix the scattering time scale at $10^{-10}$ (red lines), and $10^{-4.4}$ (blue lines) chosen to represent no scattering, and the maximum-likelihood timescale from the three-dimensional analysis.   (Bottom-right panel)  Comparison of the true posterior probability distribution for the ToA from the three-dimensional analysis compared to a Gaussian approximation, as would be performed in traditional pulsar timing analysis.}
\label{figure:MoneyPlot}
\end{figure*}

For example, \cite{2001ApJ...562L.157L} use  profile data from 4.9~GHz observations to construct a model for the intrinsic pulse profile, and then assume this model when determining the scattering time scales and scaling with frequency.  However, significant profile evolution is known to occur across wide frequency ranges which can bias these parameter estimates.  In this case the uncertainties on the measured parameters were multiplied by a factor of three, however this is clearly unsatifactory, and a more statistically robust approach would be preferred.   In \cite{2003ApJ...584..782B}, no assumptions are made about the intrinsic shape of the pulse profile,  however the DM is assumed fixed, and the \textsc{CLEAN} algorithm is sub-optimal when in the scattering timescale is small.

Further complications arise when wanting to include scattering variations in timing analysis.  Existing approaches have suggested obtaining an estimate for the scattering time scale, and then using that to `correct'  the ToAs (e.g., \citealt{2016ApJ...818..166L}). While cyclic spectroscopy has shown significant promise, the robustness of the method breaks down for low signal-to-noise profiles (S/N), where the S/N is less than
$\sim$ 100 \citep{2015ApJ...815...89P}.  Even in the high S/N simulations, however, the approach advocated is to `correct' the ToAs using estimates of the scattering delay obtained from the cyclic spectroscopy.   In the PSR J1643$-$1224 data set analysed in Section~\ref{Section:Results} the mean S/N per epoch is $\sim$ 90, with a minimum of only 15. We therefore require an approach that is robust across all observed profile S/N.  Further, neither approach accounts for the significant covariances that exist between the scattering timescales and other parameters of interest.

We illustrate two of the key challenges associated with incorporating scattering measurements into timing analysis in Fig.~\ref{figure:MoneyPlot}. We simulate a pulsar with a 4.6~ms period, using a Gaussian pulse shape with a full-width at half-maximum of 5\,\%  pulse phase.  The scattering time scale is chosen to be $10^{-4.5}$\,s at a reference frequency of 1~GHz, consistent with the values observed in our analysis of PSR J1643$-$1224 in Section~\ref{Section:Results}.  We use a frequency range of 1.25-1.5\,GHz, separated into eight channels, again chosen to be consistent with a typical 20\,cm observation used in  Section~\ref{Section:Results}.  In the top left panel we show the simulated pulse profile at the lowest-frequency channel (red lines, 1263\,MHz), and the highest-frequency channel (black lines, 1475\,MHz).  We have aligned the leading edge of the profiles (shown as a zoomed in region in the bottom sub-panel), and can see that the trailing edge has been broadened in the low-frequency channel  (shown as a zoomed in region in the top sub-panel).  In the top-right panel we show the same thing, however after adding noise to the simulation.  By eye the scattering is undetectable, however we will now show that it is still sufficient to significantly bias parameter estimates when ignored in the analysis.

In the bottom-left panel, we show in black the one- and two-dimensional marginalised posterior distributions from a three-dimensional analysis of this simulated observation.   The parameters included in the analysis are:

\begin{itemize}
 \item[(i)] the ToA, measured in units of phase, 
 \item[(ii)] the DM in units of the \textsc{Tempo2} uncertainty, 
 \item[(iii)] and $\log_{10}$ of the scattering timescale, $\tau$, measured in seconds at a reference frequency of 1GHz.  
\end{itemize} 
 
One can clearly see that the scattering timescale is correlated in a non-linear way with both the pulse ToA,  and the DM.  The ToA in particular has a bi-modal distribution, with one peak associated with `large' scattering timescales ($\log_{10} \tau \ga -5$), and one associated with small scattering time scales ($\log_{10} \tau \la -5$).  Note that the peak of the one-dimensional probability distribution for the scattering time scale is consistent with the simulated value, indicating that despite being practically invisible by eye, it is still marginally detected in our analysis.

In the same panel we show the posterior distributions for the phase and DM when fixing the scattering timescale at $10^{-10}$ (red lines), and $10^{-4.4}$ (blue lines).  These correspond to models with no scattering, or using the maximum likelihood scattering timescale obtained from the three-dimensional analysis.  In both cases this results in significant bias in both the measured values, and the uncertainties of the other two parameters.  In particular, the measured ToA is completely inconsistent between the two models, with a difference of approximately 10\,$\mu$s, two to three orders of magnitude more than the shift expected from an isotropic gravitational wave background in pulsar timing observations given the most stringent current upper limits \citep{2015Sci...349.1522S}.   While the frequency dependence of scattering variations will help to decrease their covariance with gravitational waves, and so will limit how much this bias truly impacts our sensitivity, clearly leaving such variations unmodelled in the analysis must be sub-optimal.

In Fig.~\ref{figure:MoneyPlot} (bottom-right panel) we then compare the posterior distribution for the ToA from the three-dimensional analysis (black line) to a Gaussian approximation of the posterior (red line).   Typically when forming ToAs, a Gaussian approximation is made of the true probability density function of the arrival time.  While methods such as those presented in \cite{2014MNRAS.443.3752L} and \cite{2014ApJ...790...93P}, which perform a simultaneous fit to broadband profile data in order to fit for both phase and DM can be extended to also fit for scattering, the fundamental approximation of Gaussianity implicit in the ToA forming process is invalid in regimes such as that shown in Fig.~\ref{figure:MoneyPlot}.  While the Gaussian approximation is conservative, it is sub-optimal compared to incorporating the full probability density function of the arrival time, and while in principle the full posterior distribution from this kind of analysis could be used in subsequent pulsar timing analysis, such an approach will always be an approximation to simply performing the analysis in the profile domain.

\begin{figure*}
\begin{center}$
\begin{array}{cc}
\includegraphics[trim = 50 50 20 0, clip,width=85mm]{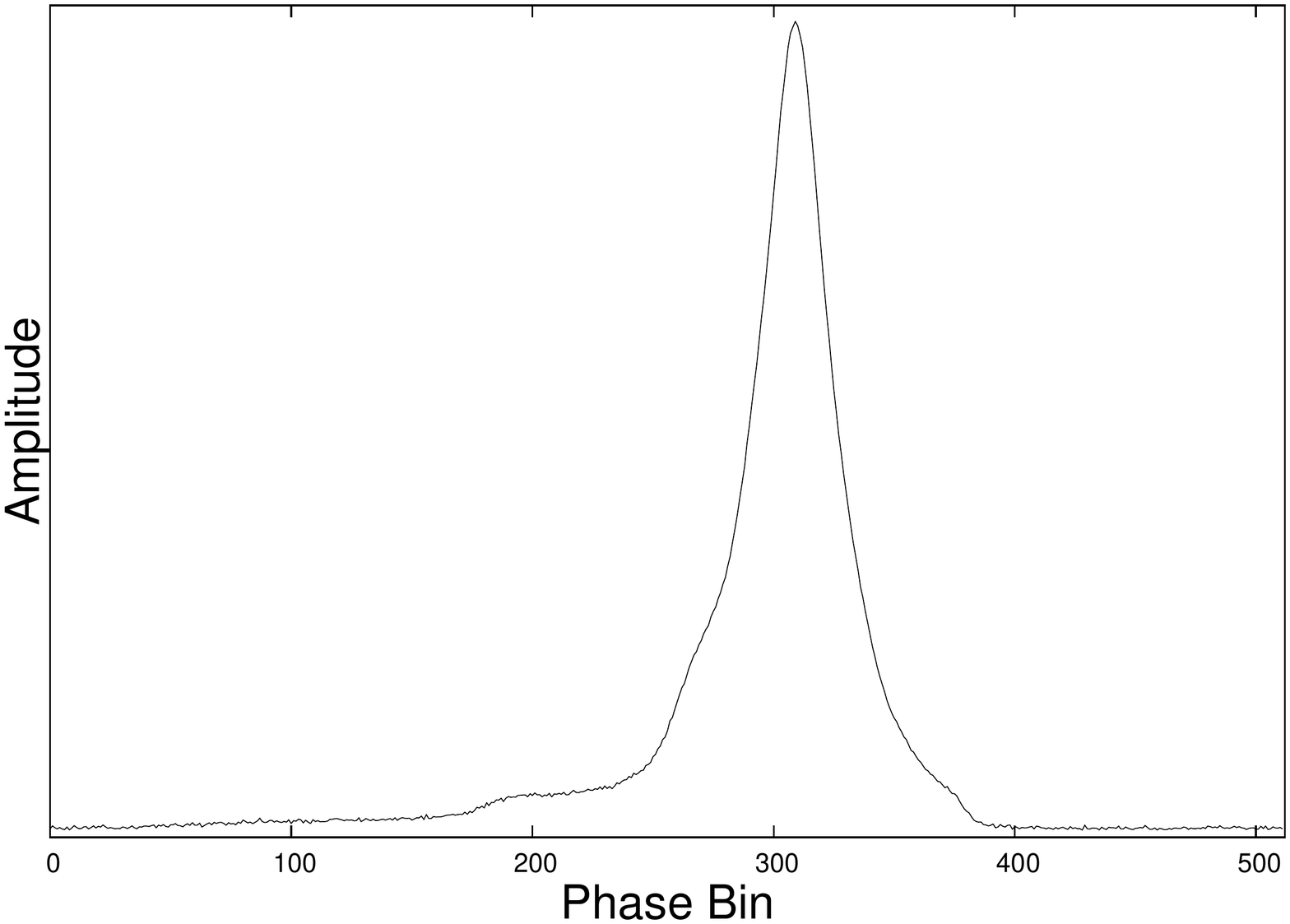} &
\includegraphics[trim = 50 50 20 0, clip,width=85mm]{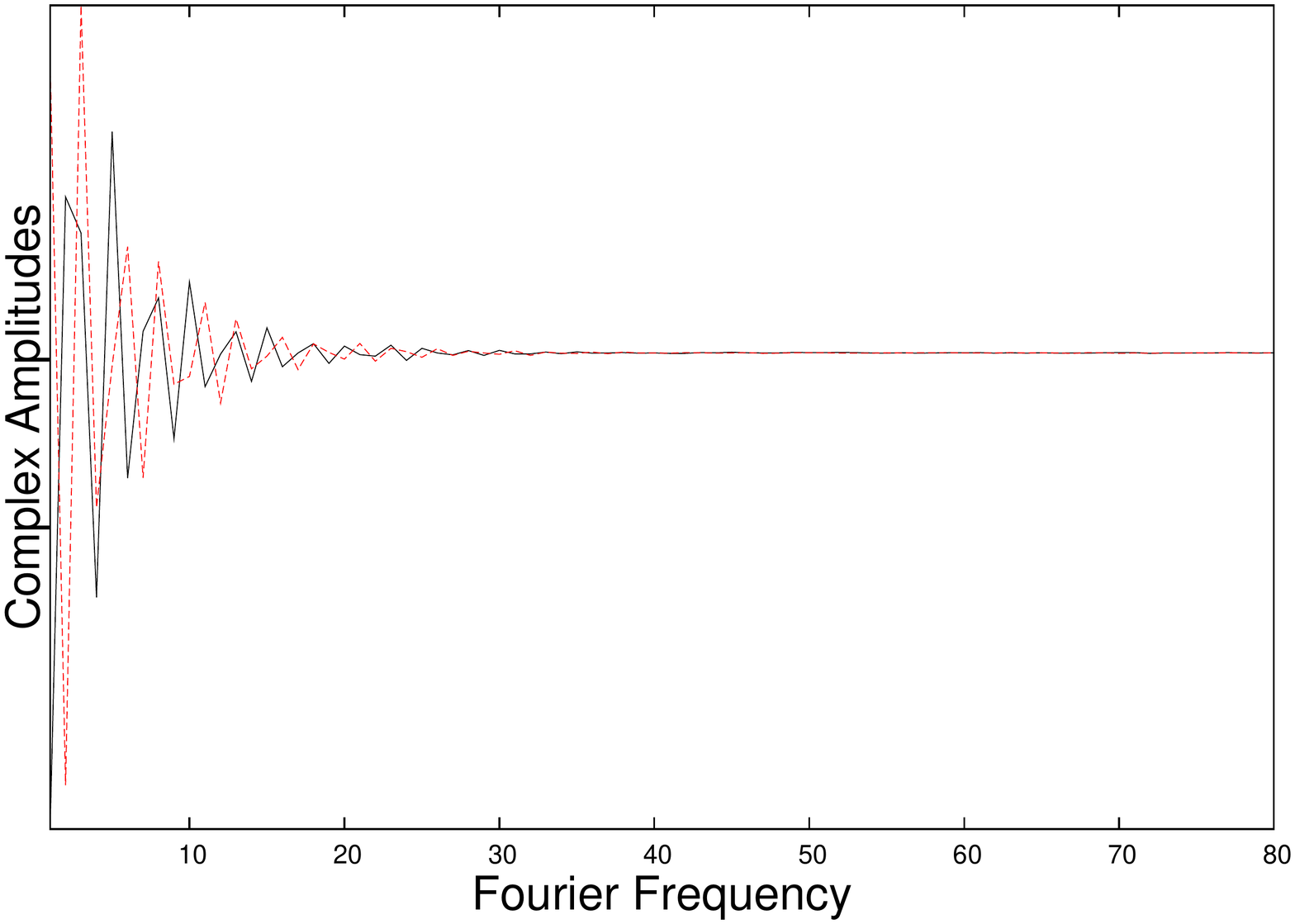}\\
\end{array}$
\end{center}
\caption{Time averaged profile for PSR J1643$-$1224, using the 1200~MHz to 1500~MHz data that we analyse in Section~\ref{Section:Results} in the time domain (left-panel), and Fourier domain (right-panel).  While the profile extends across over 300 phase bins, over 99\% of the signal is contained in the first 40 Fourier bins.}
\label{figure:ProfileTrunc}
\end{figure*}

In this paper we present a solution to these problems by extending the profile domain pulsar timing framework (\citealt{2016arXiv161205258L} and references therein, henceforth L16) to incorporate scattering variations.  This means that the parameters that describe the scattering as a function of time can be estimated simultaneously with all other parameters in the model.  This includes DM variations, the pulsar timing model, timing noise, pulse jitter, and also any models for the profile, profile evolution as a function of frequency, and pulse shape instability.  

In order to incorporate scattering into the existing framework we perform the full analysis in the Fourier domain, as opposed to the time domain as in L16.  In Section~\ref{Section:Like} we describe this modified framework, before applying it to simulated, and then real data in Sections~\ref{Section:Simulations}-\ref{Section:Results}, before ending with some concluding remarks in Section~\ref{Section:Conclusions}.  All the analysis presented subsequently is performed using a new, publicly available GPU accelerated code\footnote{https://github.com/LindleyLentati/TempoNest2}, and we have included the simulated data sets used in Section~\ref{Section:Simulations} in the repository with instructions for how to replicate the results presented in this paper.

\section{A Fourier-domain model}
\label{Section:Like}

In the following section we will build on the methods presented in L16.  In that previous work, the profile-domain analysis was performed in the time-domain.  Here, we will define our models in the Fourier-domain which allows us to optimize the analysis in two ways.  Firstly, the Fourier representation makes it trivial to truncate our model for the profile at a particular harmonic, decreasing the size of the linear-algebra operations required to evaluate it.  Secondly, when including scattering in the analysis, we only need to multiply the pulse broadening function with our profile model in the Fourier domain, as opposed to computing a convolution in the time domain, and the gradients can likewise be obtained at less computational expense for all parameters.

For a full description of the general profile domain framework we refer the reader to  L16.  Below we will give details of how the methodology has been changed to allow analysis in the Fourier domain, and of our implementation of models for scattering.

\subsection{The Profile Model}
\label{section:Shapelets}

As in L16 we construct our profile model using the shapelet basis \citep{2003MNRAS.338...35R}, for which the Fourier-transform can be obtained analytically.  In the time domain, shapelets are described by a position $t$, a scale factor $\Lambda$, and a set of $n_\mathrm{max}$ amplitude parameters, with which we can construct the set of basis functions:

\begin{equation}
B_n(t;\Lambda) \equiv \Lambda^{-1/2} \Phi_n(\Lambda^{-1}t),
\end{equation}
with $\Phi_n(t)$ given by:

\begin{equation}
\Phi_n(t) \equiv \left[2^nn!\sqrt{\pi}\;\right]^{-1/2} H_n\left(t\right)\;\exp\left(-\frac{t^2}{2}\right),
\end{equation}
where $H_n$ is the $n$th Hermite polynomial.     In the Fourier domain, the dimensionless basis functions $\Phi_n(t)$ become:

\begin{equation}
\tilde{\Phi}_n(f) = i^n \Phi_n(f)
\end{equation}
and the basis functions $B_n(t; \Lambda)$ become:

\begin{equation}
\tilde{B}_n(f; \Lambda) = i^n B_n(f; \Lambda^{-1})
\end{equation}
for Fourier frequencies $f$.  We can then write our profile model in the Fourier domain, $\tilde{s}(f, \mathbf{\zeta}, \Lambda)$,  as the sum:

\begin{equation}
\label{Eq:oldshapefunction}
\tilde{s}(f, \mathbf{\zeta}, \Lambda) = \sum_{n\mathrm{=0}}^{n_{\mathrm{max}}} \zeta_n(\nu)\tilde{B}_n(f; \Lambda^{-1}),
\end{equation}
where $\zeta_n$ are the shapelet amplitudes, and  $n_{\mathrm{max}}$ is the number of shapelet basis vectors included in the model.  

As in L16 we have explicitly written the shapelet amplitudes as a function of the observing frequency $\nu$, and use a general polynomial expansion of the shapelet amplitudes with frequency in order to model any potential smooth profile evolution.  This model is defined such that for the $p$ terms in the polynomial we can write the $n$th shapelet amplitude $\zeta_n(\nu)$ as:

\begin{equation}
\zeta_n(\nu) = \sum_{k=0}^p(\nu - \nu_c)^k\zeta_{n,k}(\nu),
\end{equation}
where $\nu_c$ is an arbitrary reference frequency, and $\zeta_{n,k}$ is the amplitude parameter for the $k$th polynomial of the $n$th term in the shapelet model.  

Lastly, as in L16, we use the shapelet basis to describe the overall profile shape, and then scale this in amplitude for each epoch.  We therefore use the zeroth-order term as a reference, taking $\zeta_0 = 1$,  leaving only $n_{\mathrm{max}}-1$  free parameters $\zeta_n$ which are the amplitudes for the shapelet components with $n>0$.  Written in this way Eq. \ref{Eq:oldshapefunction} becomes:

\begin{equation}
\label{Eq:shapefunction}
\tilde{s}(f, A, \mathbf{\zeta}, \Lambda) = A\sum_{n\mathrm{=0}}^{n_{\mathrm{max}}} \zeta_n(\nu)\tilde{B}_n(f;\Lambda),
\end{equation}
with $A$ the overall scaling factor for a particular epoch.

\subsection{Shapelet interpolation}
\label{Section:Interpolate}

As in L16 we do not re-evaluate our shapelet model for every likelihood calculation, but adopt an interpolation scheme, where the shapelet basis vectors are precomputed on a grid from $t=0$ up to the duration of the longest phase bin in the data set.   In principal, as we are performing our analysis in the Fourier domain we can rotate the shapelet model exactly by  muliplying our complex shapelet model by a rotation vector:

\begin{equation}
\label{Eq:RotVec}
R(f, \delta \phi) = \exp\left(2\pi i f \delta \phi\right),
\end{equation}
where $\delta \phi$ is the amount of phase by which to rotate the model. However, when performing the sampling with Hamiltonian Monte Carlo (HMC) methods, we must compute the gradients for all the shapelet amplitudes.  As such, if we rotate the profile model to match the data, we must also rotate all the basis vectors for each profile in the data set.  We  find it is more efficient to pre-compute sub-bin shifts in the profile model, and then to rotate the data by an integer number of bins, again using Eq. ~\ref{Eq:RotVec} so that it aligns with the nearest interpolated set of basis functions.   In this way we need only perform a single rotation per profile, which is a significant computational saving. 

As in L16, we use an interpolation interval of 1~ns, chosen to be sufficiently small that no bias enters our analysis as a result of the interpolation process.

\subsection{Truncating the Shapelet Model}

In addition to the interpolation scheme described in Section~\ref{Section:Interpolate}, the Fourier representation makes it straightforward to truncate the shapelet model at a particular harmonic.  This means that we need only include the subset of the harmonics in the profile model that contribute above some threshold in the analysis.  While the interpolated shapelet model used in L16 significantly reduced the time taken to evaluate the mean profile compared to numerical evaluation, it still required a $\mathrm{N}_\mathrm{b}\times\mathrm{N}_\mathrm{c}$ matrix-vector product, where $N_c$ is the number of amplitudes in the shapelet model, and $N_b$ is the number of bins in the profile data.  By only including N$_\mathrm{h}$ harmonics in the profile model, we can reduce this to a $2\mathrm{N}_\mathrm{h}\times\mathrm{N}_\mathrm{c}$ product, providing an immediate decrease in the computation time of the profile model by a factor of $\mathrm{N}_\mathrm{b}/2\mathrm{N}_\mathrm{h}$.    

In Fig.~\ref{figure:ProfileTrunc} we compare the time domain (left-panel), and Fourier domain (right-panel) representations of the time averaged profile for PSR J1643$-$1224, using the 1200~MHz to 1500~MHz data that we analyse in Section~\ref{Section:Results}.    While the time-domain profile extends across $\sim 300$ phase bins, the Fourier representation contains over 99\% of the total signal in only 40 harmonics, representing a significant reduction in the size of the matrix-vector multiplication required to evaluate the model.  In Section~\ref{Section:Results} we include up to the 80th harmonic in our model.  Beyond this the relative contribution of higher harmonics in the profile model is less than one part in $10^{10}$.

In principle, variations in the shape of the pulse profile may introduce fluctuations at harmonics higher than that required for the mean profile model.  In the analysis presented here, such fluctuations will be absorbed by the white noise component of our model.  In principle, however, one could compare models that incorporate these higher frequency terms and determine the optimal maximum harmonic to include in the analysis at the expense of increased computation time.

\subsection{Scattering}
\label{section:Scattering}

The primary addition to the model used in our analysis in this work, as opposed to L16, is broadening of the pulse profile due to scattering as the pulse passes through the IISM.   In the time-domain approach, pulse broadening would have to be included as a convolution of the pulse profile with a pulse broadening function (PBF).  Typically this is performed by performing a Fast Fourier Transform (FFT) of the model and the PBF, multiplying them in the Fourier-domain, and then transforming back to the time domain.   When calculating the gradient of the likelihood for each parameter in the model,  these too require Fourier-transforms to be performed, and so the complexity of the likelihood calculation grows rapidly.  In  the Fourier domain we need only perform the multiplication of the  Fourier transform of the PBF with our profile model, and the gradients can likewise be obtained trivially for all parameters.

In principal, any model for scattering that allows for the calculation of a gradient can be incorporated into our analysis.  Even if the analytic Fourier transform of the PBF is not known, if the gradient can be computed in the time domain it can then be transformed via FFT for use in our analysis framework, although this will be less computationally efficient.  In our analysis in this work we simply assume a  thin screen model for the PBF \citep{1972MNRAS.157...55W}, using a single parameter for the time scale, $\tau$, which in the time domain will be given by:

\begin{equation}
\label{Eq:TimePBF}
\mathrm{PBF}(t,\bar{\tau}, \nu, \alpha) = H(t)\exp{\left(-\frac{t}{\nu^{-\alpha}10^{\bar{\tau}}}\right)}. 
\end{equation}
where $\nu$ is the observing frequency, and $H(t)$ is the Heavyside step function.  Note that, as we always require the scattering timescale to be positive, we fit for $\log_{10}$ of the scattering timescale, $\bar{\tau}$.  The analytic Fourier transform of Eq. \ref{Eq:TimePBF} is then given by:

\begin{equation}
\label{Eq:FreqPBF}
\mathrm{PBF}(f, \bar{\tau}, \nu, \alpha) = \frac{1}{(2\pi f \nu^{-\alpha}10^{\bar{\tau}})^2 + 1} + \frac{-2\pi f \nu^{-\alpha}10^{\bar{\tau}}}{(2\pi f \nu^{-\alpha}10^{\bar{\tau}})^2 +1}i,
\end{equation}
with gradients with respect to $\bar{\tau}$ and $\alpha$:
\begin{eqnarray}
\frac{d\mathrm{PBF}}{d\bar{\tau}} &=& \frac{\log(10)}{((\omega \nu^{-\alpha}\bar{\tau})^2 +1)^2} \nonumber \\
&\times& \left(-2\omega^2 \nu^{-2\alpha} 10^{2\bar{\tau}} + \omega \nu^{-\alpha}10^{\bar{\tau}}((10^{\bar{\tau}}\omega\nu^{-\alpha})^2 -1)i\right),
\end{eqnarray}
\begin{eqnarray}
\frac{d\mathrm{PBF}}{d\alpha} &=& \frac{-\log(\nu)}{((\omega \nu^{-\alpha}\bar{\tau})^2 +1)^2} \nonumber \\
&\times& \left(-2\omega^2 \nu^{-2\alpha} 10^{2\bar{\tau}} + \omega \nu^{-\alpha}10^{\bar{\tau}}((10^{\bar{\tau}}\omega\nu^{-\alpha})^2 -1)i\right),
\end{eqnarray}
where as before  $i$ indicates a complex number and $\omega = 2\pi f$.  We can then multiply the Fourier representation of our shapelet model, $\mathbf{\tilde{s}}$, by Eq.~\ref{Eq:FreqPBF} to get the scattered profile model.

In our analysis of both simulated data, and the PSR J1643 data set, we use a piecewise-constant $\tau$(t) model for the scattering.  This therefore makes the same assumptions made when modelling DM variations using the DMX parameterisation \citep{2013ApJ...762...94D}.  In principal, one could also model the power spectrum of the scattering, under the assumption that it is a smooth, wide-sense stationary process in the same manner as the smooth model for DM variations in L16.  Deviations from this simple model, such as those predicted from simulations \citep{2010ApJ...717.1206C} could then be encapsulated as additional, non-stationary features in the model, in the same manner as the extreme scattering events observed in PSR J1713+0747 (e.g. \citealt{2016MNRAS.458.3341D, 2016MNRAS.458.2161L}). However, for this demonstration of the profile domain approach, we do not yet consider such a model.

\subsection{Evaluating the profile domain model}

The remainder of the timing framework described in L16 requires only minimal changes to operate on the Fourier representation of the profile data.  As in L16, we consider the data in terms of a set of $N_\mathrm{e}$ epochs. Each epoch $i$ then has $N_{\mathrm{c},i}$ channels, such that the total number of profiles  $N_\mathrm{p} = \sum_{i=1}^{N_\mathrm{e}} N_{\mathrm{c},i}$.  The profile in the $j$th channel of the $i$th epoch then consists of a set of $N_{i,j}$ values representing the amplitude of the profile as measured at a set of times $\mathbf{t_{i,j}}$ which we denote $\mathbf{d}_{i,j}$.  The Fourier representation of the profile data, $\mathbf{\tilde{d}}_{i,j}$ will therefore be a set of $N_{i,j}/2 +1$ complex values.  While in L16 a baseline offset for each profile was required as part of the model, in the Fourier representation we simply discard both the DC offset term and the nyquist term, so that our final data vector for each profile is of length $N_{i,j}/2 - 1$.   

All further parameters, such as those describing pulse jitter or shape variation, can then  be incorporated into our Fourier-domain shapelet model $\tilde{s}$ in exactly the same manner as in L16.  We can then write the final likelihood that the data is described by the parameters in our model, which we collectively refer to as  $\theta$, as:

\begin{eqnarray}
\label{Eq:TimeLike}
\mathrm{Pr}(\mathbf{\tilde{d}} | \mathbf{\theta}) &\propto& \prod_{i=1}^{N_\mathrm{e}}\prod_{j=1}^{N_{\mathrm{c},i}} \frac{1}{\sqrt{\mathrm{det}\mathbf{N_{i,j}}}}\\
&\times& \exp{\left[-\frac{1}{2}(\mathbf{\tilde{d}_{i,j}} - \mathbf{\tilde{s}_{i,j}})^T\mathbfss{N}_\mathbf{{i,j}}^{-1}(\mathbf{\tilde{d}_{i,j}} - \mathbf{\tilde{s}_{i,j}})\right]} \nonumber,
\end{eqnarray}
where $\mathbfss{N}_\mathbf{{i,j}}$  is the white noise covariance matrix for the Fourier domain profile corresponding to the $j$th channel in the $i$th epoch, with elements $(N_{i,j})_{mn} = \sigma_{i,j}\delta_{mn}$, with $\sigma_{i,j}$ the root-mean-square (RMS) deviation of the uncorrelated radiometer noise in the profile.  When performing the analysis with the Guided Hamiltonian Sampler (GHS), $\sigma_{i,j}$ is a free parameter in our analysis for every profile.

\section{Application to Simulated Data}
\label{Section:Simulations}

\begin{figure*}
\begin{center}$
\begin{array}{cc}
\includegraphics[trim = 40 10 0 0, clip,width=80mm]{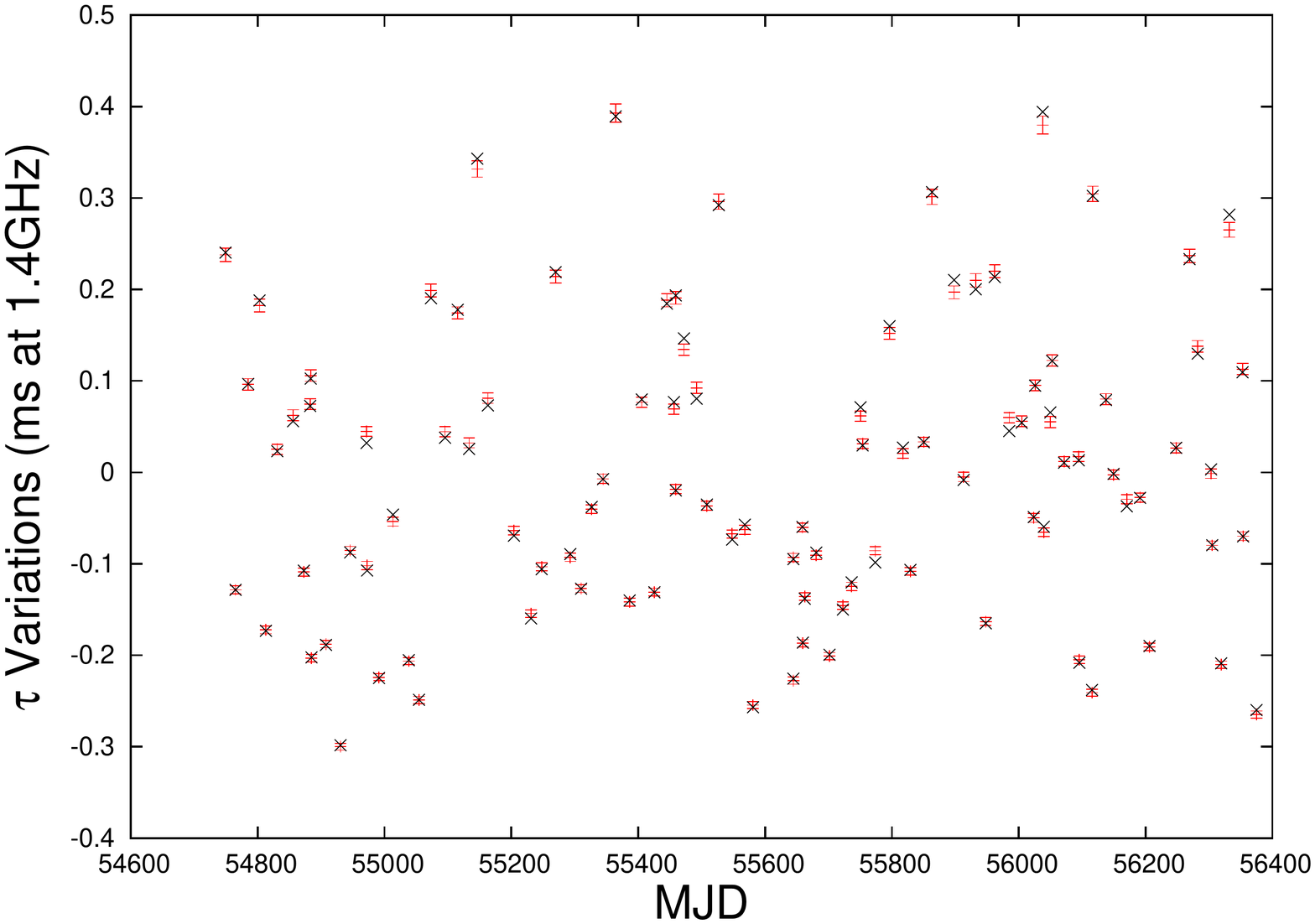} &
\includegraphics[trim = 40 0 0 0, clip,width=80mm]{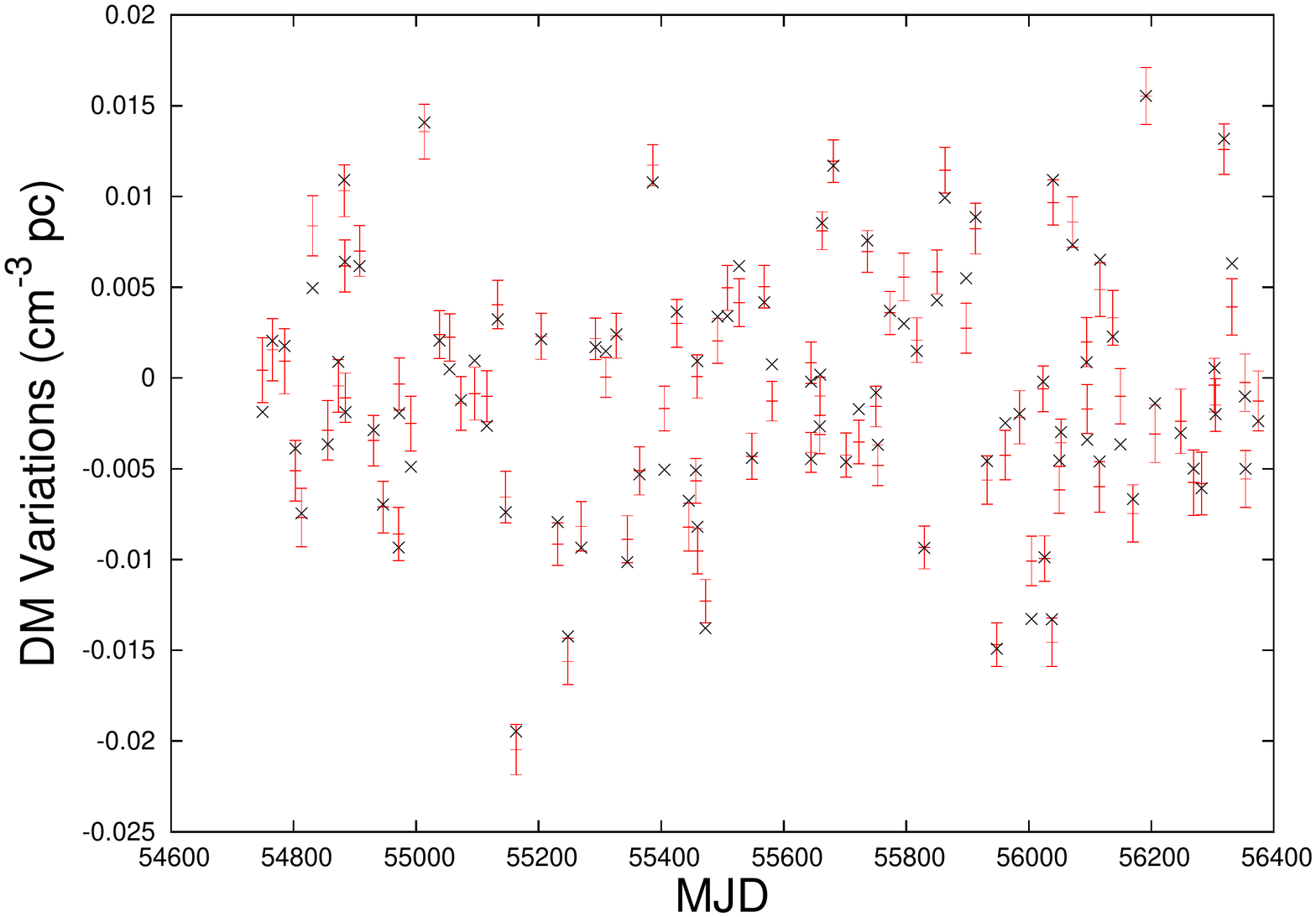}\\
\includegraphics[trim = 40 10 0 0, clip,width=80mm]{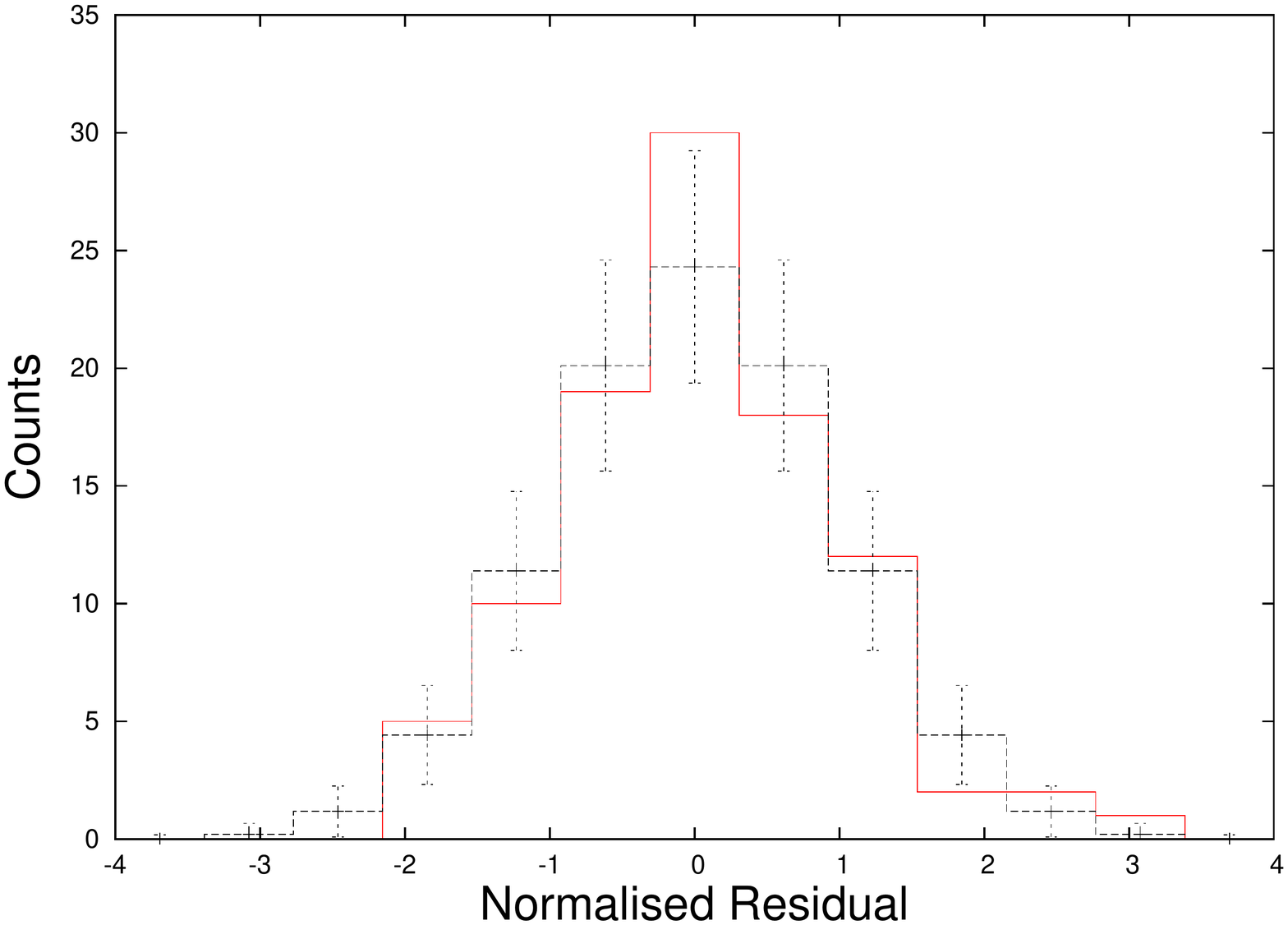} &
\includegraphics[trim = 40 0 0 0, clip,width=80mm]{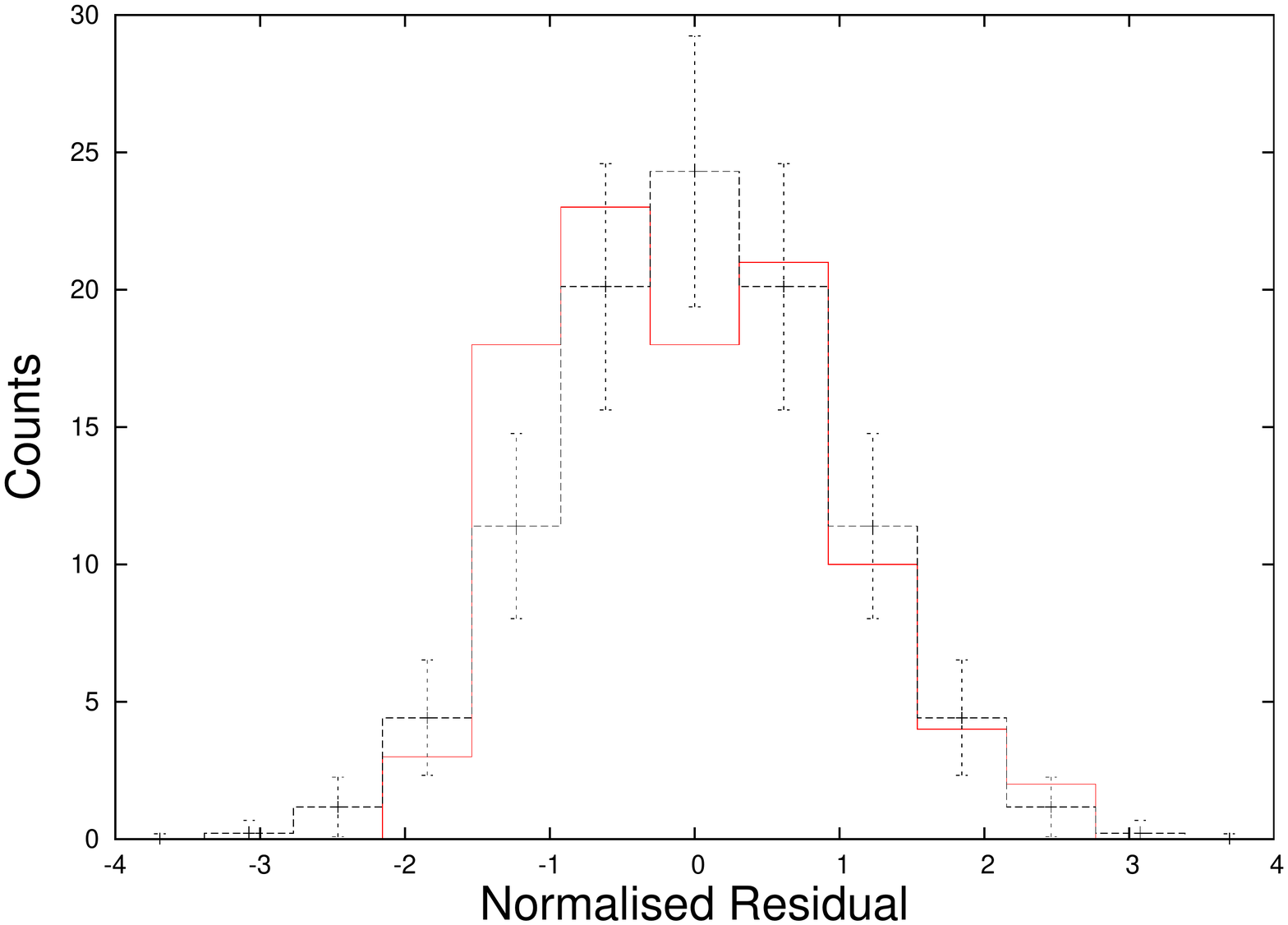}\\
\includegraphics[trim = 40 0 0 0, clip,width=80mm]{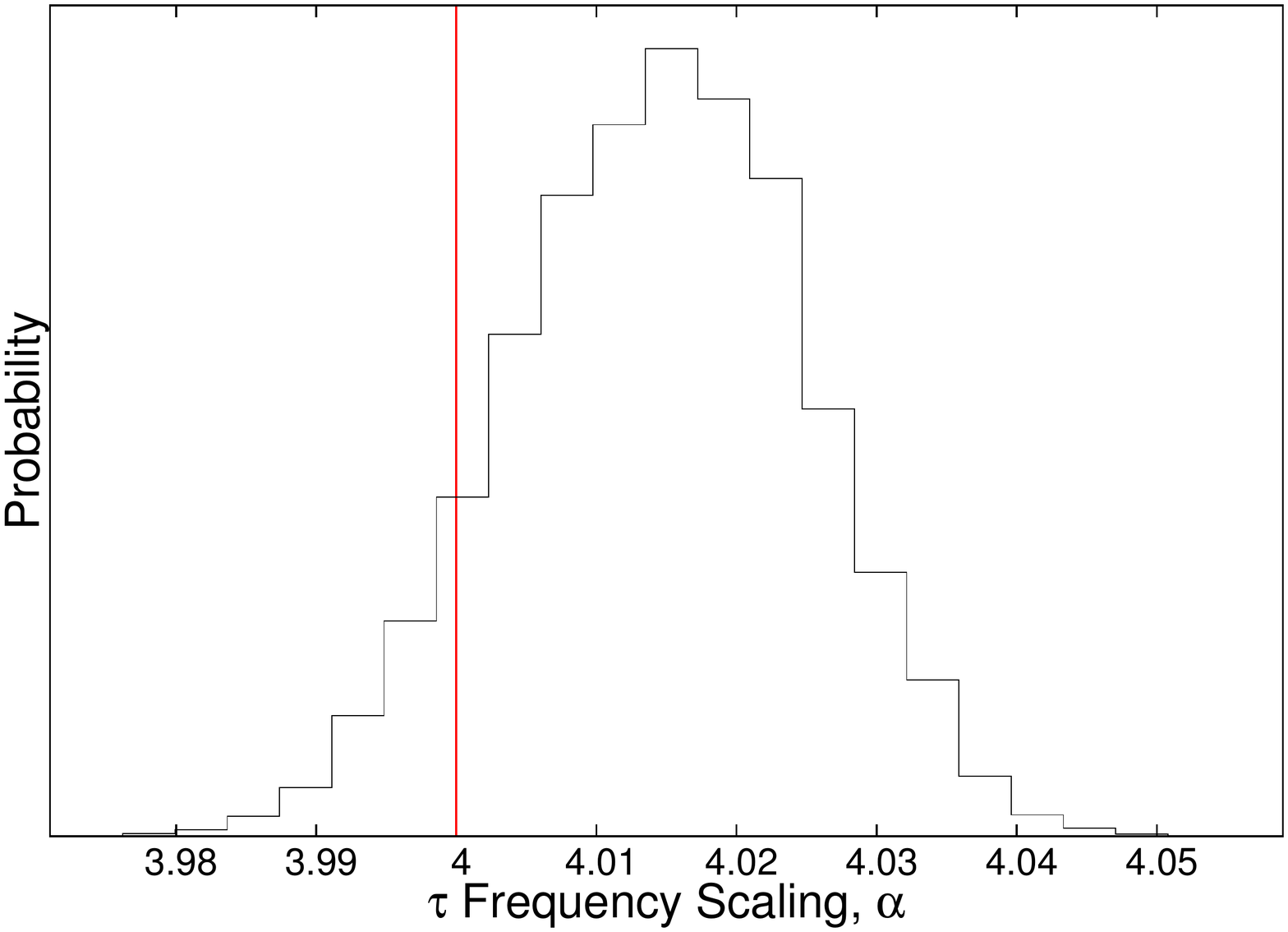} &
\includegraphics[trim = 40 0 0 0, clip,width=80mm]{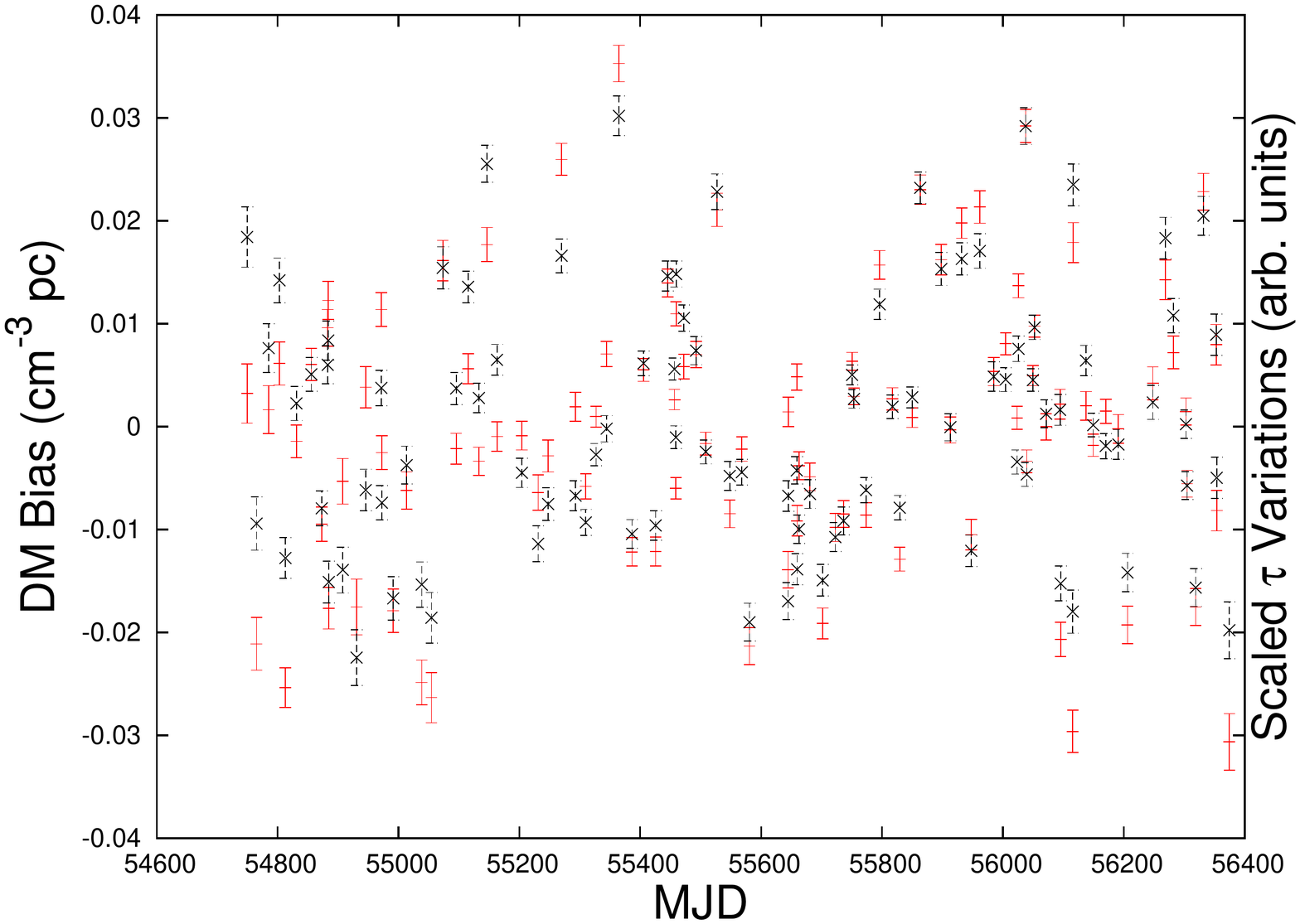} \\
\end{array}$
\end{center}
\caption{/cm{(Top panels) Injected values (red crosses) for scattering (left panel) and DM variations (right panel) used in the simulated data set analysed in Section~\ref{Section:Simulations}, and the recovered parameter estimates and 1~$\sigma$ confidence intervals from the global analysis (black points with uncertainties).  (Middle panels) Histograms of the residuals (red lines) after subtracting the maximum likelihood parameter estimates from the injected values, compared to a Gaussian distribution with unit variance (black lines) for the scattering (left panel) and DM variations (right panel).  Error bars represent $\sqrt{N}$ uncertainties on the expected Gaussian distribution.  (Bottom left) One-dimensional marginalised posterior probability distribution for the frequency scaling parameter, $\alpha$ (c.f., Eq.~\ref{Eq:FreqPBF}) from our analysis of the simulated data set described in Section~\ref{Section:Simulations}.  The simulated value is indicated with the vertical red line. (Bottom right)  Difference between the measured DM variations from the simulated data set when including scattering variations in the model, or when assuming only a mean scattering parameter (red points, 1$\sigma$ uncertainties).  We compare this with a shifted and scaled measurement of the scattering variations (black points, 1$\sigma$ uncertainties).  The Pearson product-moment correlation coefficient between the red and black points is $0.92\pm 0.04$. }}
\label{figure:SimPlot}
\end{figure*}

In order to test the efficacy of the analysis method described in the preceding sections, we first apply it to a simulated data set.  In particular we simulate 100 observational epochs, each assuming 256~MHz of bandwidth split into 8~channels, with a central frequency of 1369~MHz and covering a total time span of 4.5 years.  Each epoch has an integrated S/N across the observing band of $\sim 500$.  Both the MJD and the bandwidth of each simulated epoch are taken from the 1400\,MHz data used in Section~\ref{Section:Results}.  We use a non-evolving, Gaussian profile, and a timing model consistent with that given in Table~\ref{Table:1643Params} for PSR~J1643$-$1224.  

In addition to this timing model, each epoch is subject to a random change in the DM with a standard deviation for the variations of $0.007$cm$^{-3}$pc, and changes in the scattering of $0.81 \pm 0.16$~ms at a reference frequency of 1.4~GHz.  The scattering is simulated using the thin screen model described in Section~\ref{section:Scattering}, with the  frequency scaling parameter, $\alpha$, equal to four.  We then perform a simultaneous analysis of all epochs, fitting for:

\begin{itemize}
 \item[i)] The DM and scattering time scale at each epoch, 
 \item[ii)] the frequency scaling of the scattering, $\alpha$,  
 \item[iii)] the timing model parameters, and 
 \item[iv)]   the pulse amplitude and instrumental noise level in each profile.
 \end{itemize}
In Fig.~\ref{figure:SimPlot} we show both the injected values (red crosses) and mean parameter estimates with 1~$\sigma$ confidence intervals obtained from this analysis (black points with error bars) for the scattering (top-left panel) and DM variations (top-right panel).  We find the parameter estimates are consistent in both cases, and that the histogram of the residuals after subtracting the maximum likelihood values are consistent with a Gaussian distribution with unit variance (middle panels, red and black lines respectively).  The Pearson product-moment correlation coefficient between the simulated and recovered values is $0.999 \pm 0.005$.   In Fig.~\ref{figure:SimPlot} (bottom-left panel) we then show the one dimensional marginalised posterior distribution for the frequency dependence of the scattering model, $\alpha$.  We indicate the simulated value with a vertical red line, and find our inferred values are consistent with the simulation.

We also use this simulated data set to check the impact of not modeling the scattering variations on the parameter estimates for the DM variations.  In  Fig.~\ref{figure:SimPlot} (bottom-right panel) we plot the difference between the measured DM variations from the simulated data set when including scattering variations in the model, or when assuming only a mean scattering timescale (red points).  Not modeling the variations in the scattering leads to significant bias in the measured DM variations. Note the different scales on the y-axis of the bottom-right, and top-right panel, the bias introduced in this case is a factor two larger than the DM variations themselves. We compare this bias with the structure present in the scattering variations (a shifted and scaled version of the scattering variations is overplotted in black).  The Pearson product-moment correlation coefficient between the two sets of measurements is $0.92\pm 0.04$, indicating significant correlation between the unmodeled scattering, and the resultant bias in the DM variations.

\section{Application to Real Data}
\label{Section:Results}

\begin{table*}
\caption{Timing Parameters for PSR J1643$-$1224}
\begin{tabular}{llll}
\hline\hline
\multicolumn{4}{c}{Measured Quantities} \\
\hline
Scenario & S1 & S2 & S3 \\
\hline
Right ascension, (hh:mm:ss)\dotfill &  16:43:38.161512(21) &  16:43:38.161572(11) &  16:43:38.161510(12)\\
Declination, (dd:mm:ss)\dotfill & $-$12:24:58.6724(14) & $-$12:24:58.6725(8)  & $-$12:24:58.6734(9) \\
Pulse frequency, (s$^{-1}$)\dotfill & 216.373337142644(5) & 216.373337142639 (3) & 216.373337142639(4)\\
First derivative of pulse frequency, (s$^{-2}$)\dotfill & $-$8.6450(5)$\times 10^{-16}$ &  $-$8.6448(3)$\times 10^{-16}$  & $-$8.6450(4)$\times 10^{-16}$ \\
Proper motion in right ascension,  (mas\,yr$^{-1}$)\dotfill & 6.00(10) & 5.50(6) & 5.85(8)\\
Proper motion in declination,  (mas\,yr$^{-1}$)\dotfill & 3.8(5) & 3.5(3)  & 3.9(4) \\
Parallax, $\pi$ (mas)\dotfill & 1.3(4) & 1.3(3)  & 1.0(3) \\
Orbital period, $P_b$ (d)\dotfill & 147.01739774(6) & 147.01739775(4) & 147.01739772(5) \\
Epoch of periastron, $T_0$ (MJD)\dotfill & 49283.9336(5) & 49283.9337(4) & 49283.9336(5) \\
Projected semi-major axis of orbit, $x$ (lt-s)\dotfill & 25.0726165(29) & 25.0726182(19) & 25.0726165(23)\\
Longitude of periastron, $\omega_0$ (deg)\dotfill & 321.8487(13) & 321.8489(10) & 321.8485(11) \\
Orbital eccentricity, $e$\dotfill & 5.05755(12)$\times 10^{-4}$  & 5.05757(10)$\times 10^{-4}$  & 5.05752(11)$\times 10^{-4}$ \\
First derivative of $x$, $\dot{x}$ ($10^{-12}$)\dotfill & $-$5.2(5)$\times 10^{-14}$  & $-$5.5(3)$\times 10^{-14}$  & $-$5.2(4)$\times 10^{-14}$ \\
\hline
\multicolumn{4}{c}{Set Quantities} \\
\hline
Epoch of frequency determination (MJD)\dotfill & 55000 & 55000  & 55000 \\
Epoch of position determination (MJD)\dotfill & 55000  & 55000  & 55000\\
Epoch of DM determination (MJD)\dotfill & 55000  & 55000  & 55000 \\
\hline
\end{tabular}\label{Table:1643Params}
\end{table*}

We perform our analysis using observations of PSR J1643$-$1224 made with the 64-m Parkes radio telescope.  We choose to analyse this particular MSP because it is known to display significant scattering variability, which has been attributed to its location behind an HII region \citep{2001PASP..113.1326G, 2005A&A...442..263V}.  As such we stress that the results presented will likely not be typical for a standard MSP. However, the benefit for more typical systems will naturally depend on combinations of factors such as the observing frequency of the data set, and the brightness of the pulsar at low observing frequencies.  Further application of the techniques we have discussed in the preceding sections to a broader, and more typical population of pulsars will occur in subsequent work. Data were collected using two receiver packages, a co-axial system at 10~cm and 40~cm, and the centre pixel of a multi-beam 20\,cm system.  Data at 3100\,MHz (co-axial, hereafter `10~cm') and 1369\,MHz (multi-beam, hereafter `20~cm') were recorded using a digital polyphase filterbank (PDFB4) with a typical resolution of 1024 channels and 1024 bins and respective bandwidths of 1024\,MHz and 256\,MHz.  Data from the lower co-axial band, centred at 732\,MHz (hereafter `40~cm'), were recorded over a 64\,MHz bandwidth with a similar polyphase system, PDFB3. More details about the observing systems and data reduction process are given in \cite{2013PASA...30...17M}.  For all three systems we average the profile data into eight channels per band.

As there will be discrete time offsets (known as jumps) between the different observing systems we include these as free parameters in our analysis together with the rest of the timing model parameters given in Table~\ref{Table:1643Params}.  In all the analysis that follows we include a model for the profile and smooth profile evolution as a function of frequency using a quadratic polynomial in the shapelet amplitudes, which we found was sufficient to describe the observed variation.  In addition,  we then include DM variations using the DMX parameterisation \citep{2013ApJ...762...94D} in which the variations in DM are modelled as a piecewise constant function,  and we use an epoch length of 30 days, chosen so that the median number of 40~cm observation per epoch is one.   Below we summarise three scenarios used to investigate the impact scattering variations have on the parameter estimates and uncertainties for the DM variations and timing model parameters.

\begin{itemize}
\item[(S1)] We include only the 10~cm and 20~cm data.   No scattering variations are included in the model. \\
\item[(S2)] As (S1) however we also include the 40~cm data. \\
\item[(S3)] As (S2), however we include scattering variations with the model described in Section~\ref{section:Scattering}, with the same cadence as the DMX parameters. 
\end{itemize}
As in our analysis of simulated data in Section~\ref{Section:Simulations}, we also attempt to recover the scaling of the scattering timescale with frequency.  However, we find that the relatively narrow 40~cm band, combined with the fact that the scattering variations are already sub-dominant to the DM variations in the 20\,cm band means we are unable to constrain the quantity in our analysis.   We find that the only quantity that varies significantly when changing the scaling factor across a range of 3-5 is the mean scattering timescale, however the variations in scattering, and their impact on the other parameter estimates are consistent throughout.  We have therefore set $\alpha=4$ in all the results discussed below, and will refer only to the magnitude of variations in the scattering timescale, rather than the absolute value.

\begin{figure*}
\begin{center}$
\begin{array}{c}
\includegraphics[trim = 50 20 20 300, clip,width=150mm]{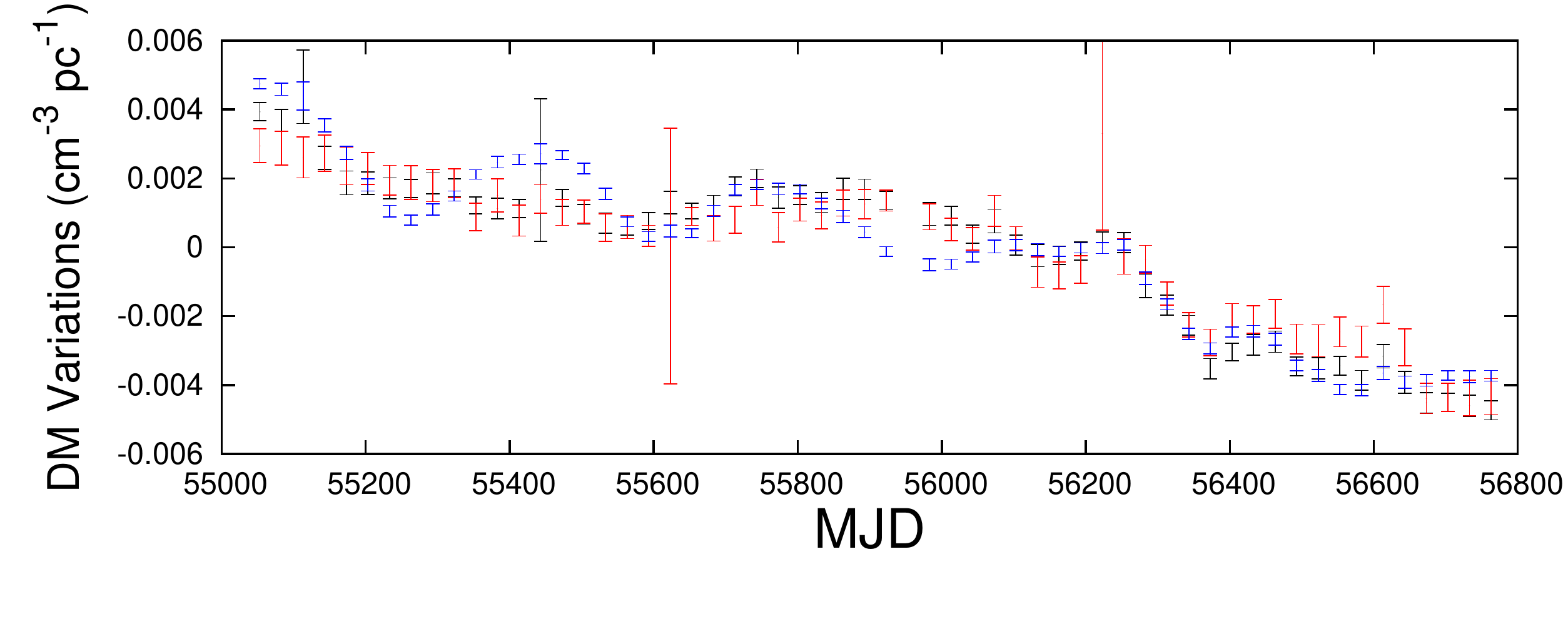} \\
\includegraphics[trim = 70 20 20 300, clip,width=150mm]{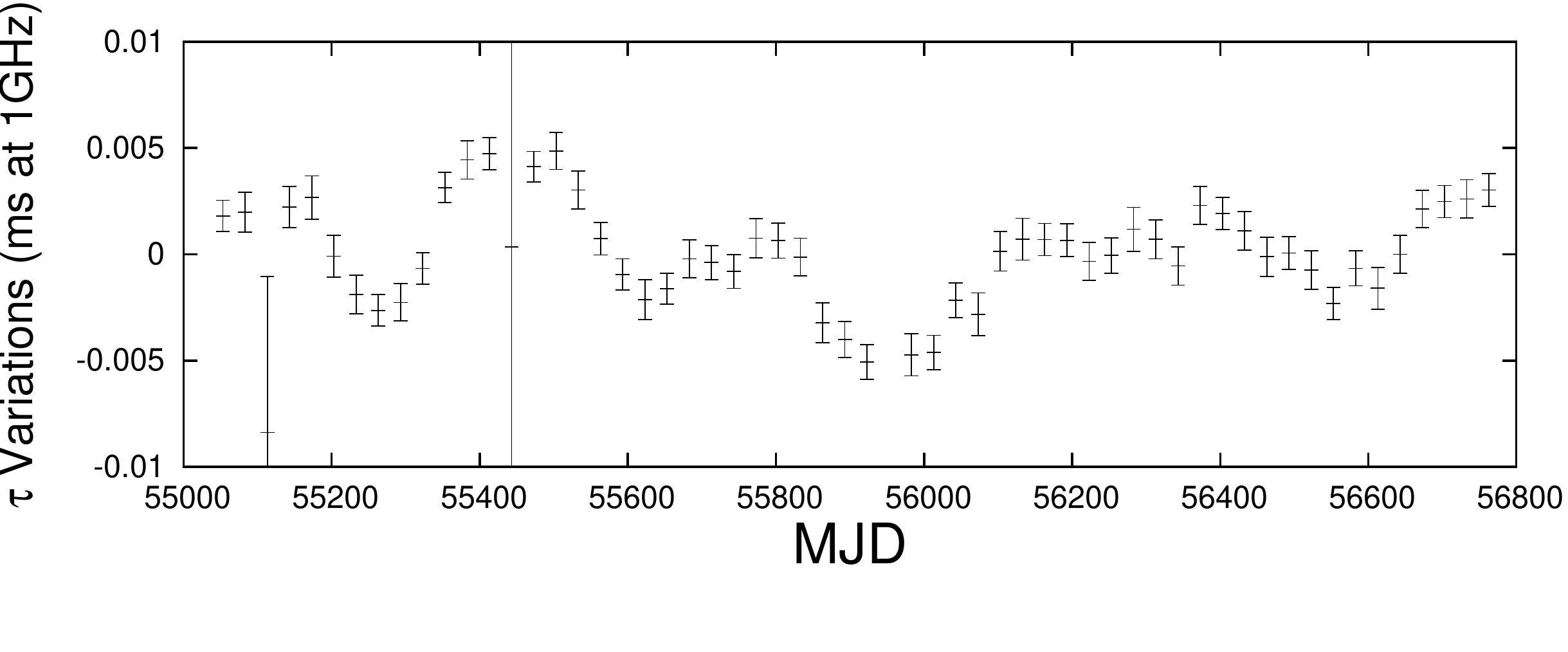}\\
\includegraphics[trim = 50 20 20 300, clip,width=150mm]{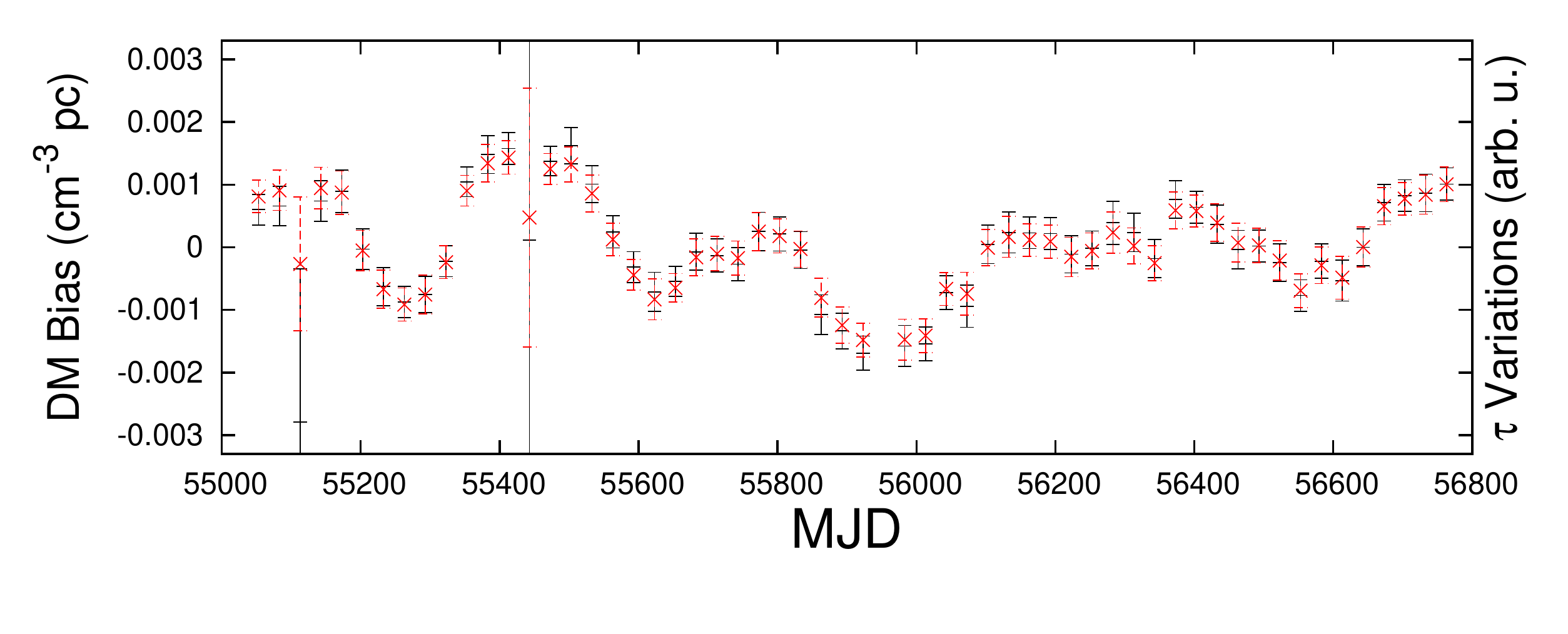}\\
\end{array}$
\end{center}
\caption{(Top Panel) Mean parameter estimates and 1~$\sigma$ confidence intervals for the DM variations in the PSR J1643$-$1224 data set for three different scenarios:  i) Including only the 10~cm and 20~cm data, and including no scattering variations in the model (red points), ii)  Including 10~cm, 20~cm and 40~cm data, and including no scattering variations in the model (blue points), iii) Including 10~cm, 20~cm and 40~cm data, and including scattering variations in the model (black points).  (Middle Panel) The variations in the scattering timescale, $\tau$, and 1~$\sigma$ confidence intervals from our analysis of the full  PSR J1643$-$1224 data set (black points).  (Bottom Panel)  Difference between the measured DM variations from scenarios~(2) and ~(3) (red points), compared to a shifted and scaled version of the scattering variations from scenario~(3) (black points).}
\label{figure:J1643DMScatter}
\end{figure*}

\subsection{DM Variations}

In Fig.~\ref{figure:J1643DMScatter} (top panel) we compare parameter estimates for the DM variations in our PSR J1643$-$1224 data set for scenarios S1-S3  (red, blue and black points respectively).

Excess time-correlated noise has been observed in this pulsar previously at low observing frequencies ($<$ 1~GHz)  (e.g., \citealt{2013MNRAS.429.2161K, 2016MNRAS.458.2161L} henceforth L16a).   In particular, L16a analysed the first International Pulsar Timing Array data set for PSR J1643$-$1224 and found that this `band-noise' had a steep dependence on observing frequency, such that above 1~GHz the timing fluctuations were consistent with DM variations only.  We can therefore expect that in scenario~(S1), we will also be dominated by DM variations, rather than scattering variations.  We find that the changes in the DM are consistent with the  smooth variations observed in L16a when including band-noise terms.

When including  the 40~cm data in our analysis in scenario~(S2), significant additional structure can be seen in the DM  that is inconsistent with the variations observed from scenario~(S1).  This structure has been reported previously in the literature as being the result of yearly variations in the DM (e.g., \citealt{2015ApJ...813...65T, 2016arXiv161203187J})

In scenario~(S3) the additional structure seen in scenario~(S2) is no longer present, and we find the parameter estimates are consistent with those obtained from scenario~(S1).   That the parameter estimates for the DM variations are inconsistent between Scenarios~(S1) and~(S2) is highly suggestive that unmodelled effects present in the 40\,cm data are introducing bias into our analysis, and are subsequently mitigated when including scattering variations into the model.

\begin{figure}
\begin{center}$
\begin{array}{c}
\includegraphics[trim = 50 40 20 60, clip,width=85mm]{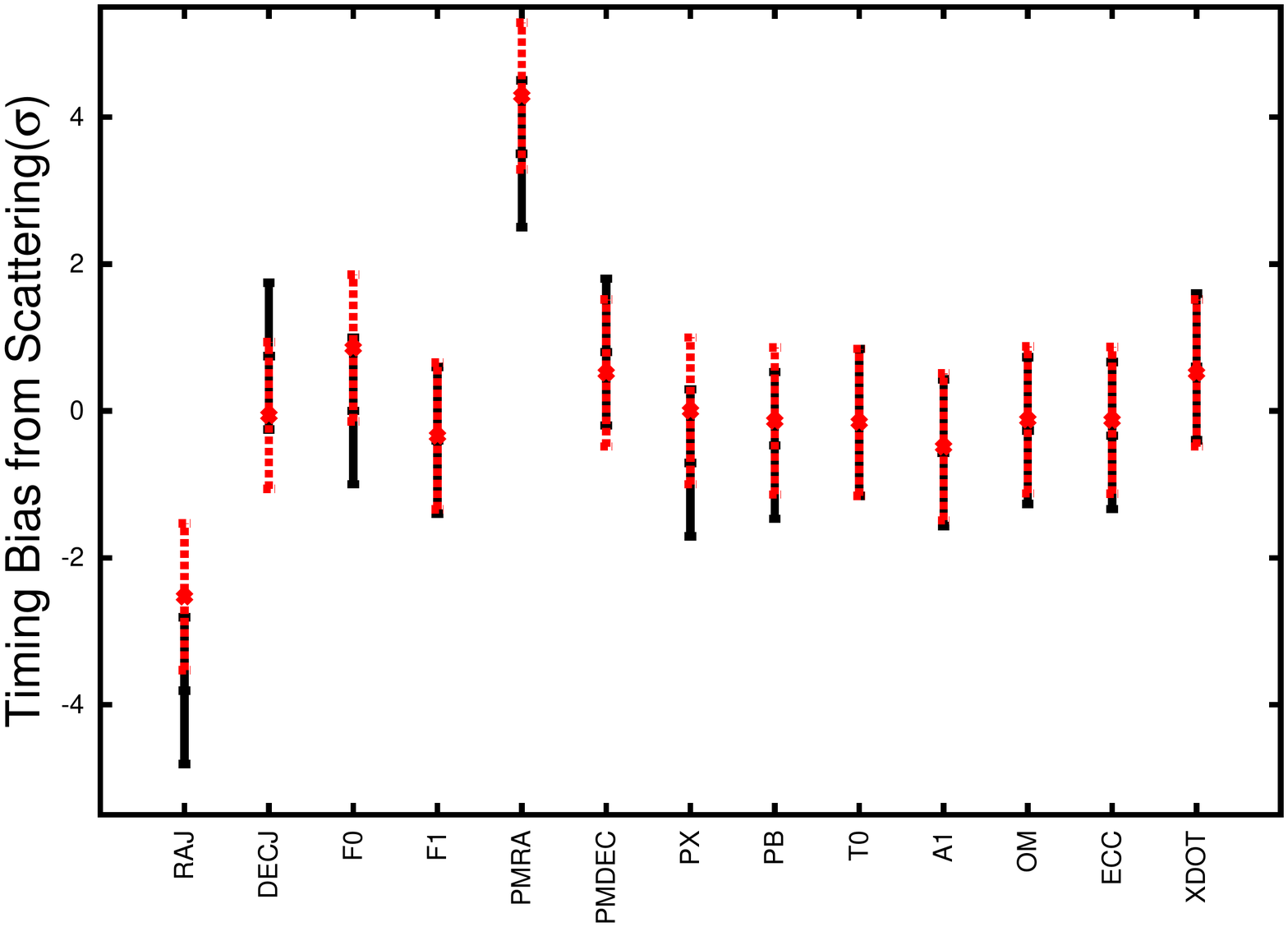} 
\end{array}$
\end{center}
\caption{Difference in the measured timing parameters in terms of their uncertainties between scenarios~(S1) and~(S2) (red points), and scenarios~(S3) and~(S2) (black points).}
\label{figure:J1643TimingBias}
\end{figure}

\subsection{Scattering Variations}

In the middle panel of Fig.~\ref{figure:J1643DMScatter} we show the parameter estimates for the variations in the scattering timescale, $\tau$, from scenario~(S3).   In the bottom panel of Fig.~\ref{figure:J1643DMScatter}  we compare this structure with the  additional structure seen in the DM variations in scenario~(S2) compared to scenario~(S1).   In particular, we overplot the difference in the DM variations from scenarios~(S1) and~(S2) (red points) with a shifted and scaled version of the scattering variations (black points).  The two clearly mirror one another, implying that the additional DM variations observed when adding low frequency data in scenario~(S2) are the result of unmodelled scattering variations, consistent with the result of not modeling the scattering variations observed in the analysis of simulated data in Section~\ref{Section:Simulations}. 

It is clear, then, that simply adding low-frequency data without considering scattering can significantly bias estimates of the DM variations.  With LOFAR \citep{2011A&A...530A..80S} already performing pulsar timing at frequencies of 100~MHz to 200~MHz, and  upcoming telescopes such as the Square Kilometre Array  providing unprecedented sensitivity in the frequency range of 500~MHz to 800~MHz, this bias introduced into the analysis when failing to appropriately model scattering variations will become extremely important, especially in the context of gravitational wave astronomy using a pulsar timing array. Looking forward this will become increasingly true even if we only consider more 'typical' pulsars compared to PSR J1643$-$1224.   This is both because in the SKA era we can expect to add many distant pulsars to PTAs, for which the presence of detectable variations in scattering will be increasingly probable, and also simply because by observing for longer the impact of such effects will grow.

\subsection{Timing Bias}

Existing methods for incorporating scattering variations into pulsar timing have suggested using maximum-likelihood estimates of scattering delays and `correcting' the measured ToA by fixing that delay in the subsequent timing analysis.  This can be done either by obtaining estimates of the delay from measurements of scintillation, as in  \cite{2016ApJ...818..166L}, or using cyclic spectroscopy (e.g., \citealt{2015ApJ...815...89P}).  With the profile domain framework we can directly compare the impact of fixing the scattering timescales in the analysis at their maximum-likelihood values, which is equivalent to simply fixing the time delay induced by the scattering when performing the analysis using ToAs.   We find that fixing the scattering timescales results in significant bias in the uncertainties for the DM variations of a factor of $\sim 1.8$, with some epochs being much larger when less multi-frequency information was available.  

In Table~\ref{Table:1643Params} we list the timing model parameter estimates for all three scenarios, and in Fig.~\ref{figure:J1643TimingBias} we then compare the parameter estimates between scenarios~(S1) and~(S2), and scenarios~(S3) and~(S2).  The bias in the parameter estimates introduced as a result of not modeling the scattering variations is substantial in several of the astrometric parameters.  From the middle panel of Fig.~\ref{figure:J1643DMScatter}  there is a clear annual term in the scattering signal which we would expect to correlate strongly with both position and proper motion.   This is indeed the case, with both position in RA, and proper motion in RA suffering  $\sim 4~\sigma$ changes in the parameter estimates.  For example, the proper motion in RA changes from $6.0\pm0.1$ to $5.50\pm 0.06$ when first adding in the 40\,cm data without a model for scattering variations, and then changes to $5.85\pm 0.08$ when including scattering variations in the model. 
%Other recently published values values for proper motion for this pulsar  6.2(1),  6.04(4), and 5.94(5) \citep{2016ApJ...818...92M, 2016MNRAS.458.3341D, 2016MNRAS.455.1751R}.

A natural question to ask is whether it is worth including low-frequency pulsar timing data into the analysis, if one is interested only in the timing model parameters and not the behavior of the ISM.  From Table~\ref{Table:1643Params} we can see that the uncertainties on the parameters from Scenario~(3) are still improved compared to Scenario~(1), in which only the 10\,cm and 20\,cm data were included in the analysis.  Therefore, if the scattering variations are modeled appropriately such that no bias is introduced in the parameter estimates, the addition of the 40\,cm data does improve the precision of the results.

\begin{figure*}
\begin{center}$
\begin{array}{c}
\includegraphics[trim = 50 20 20 300, clip,width=150mm]{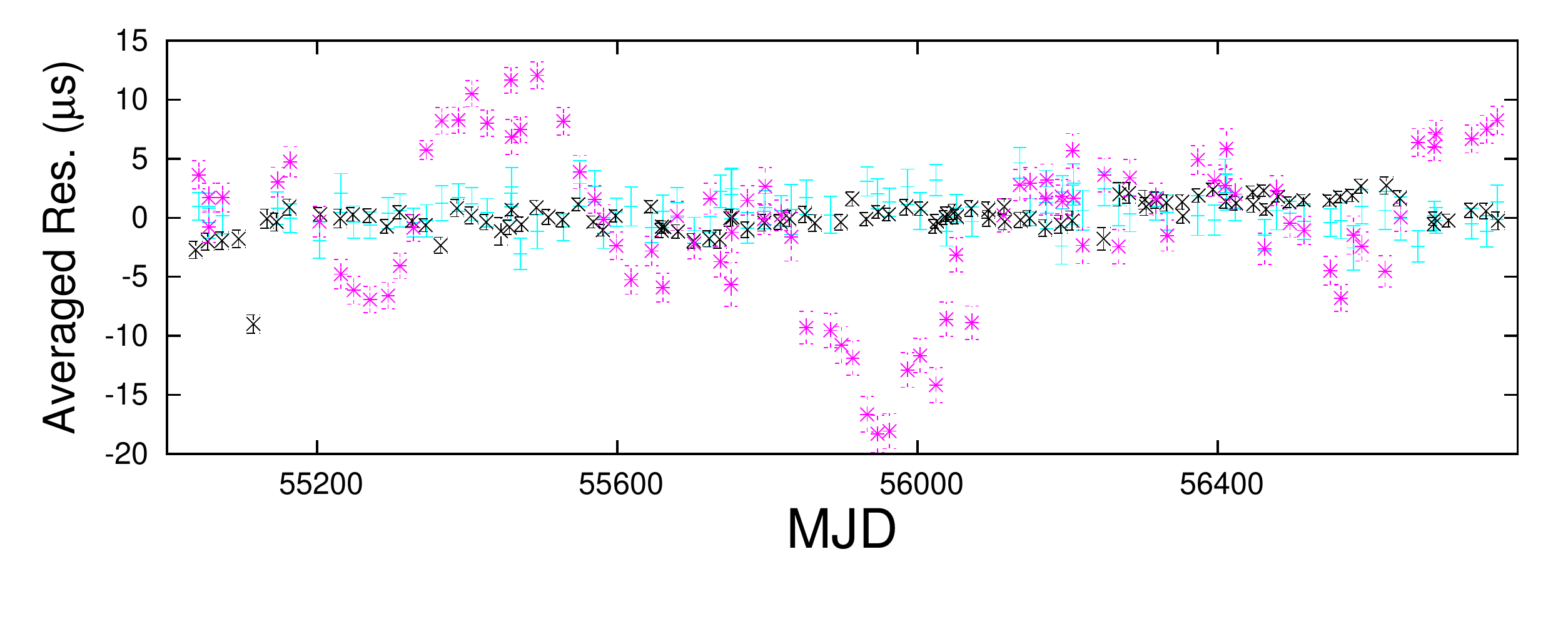} \\
\includegraphics[trim = 50 20 20 300, clip,width=150mm]{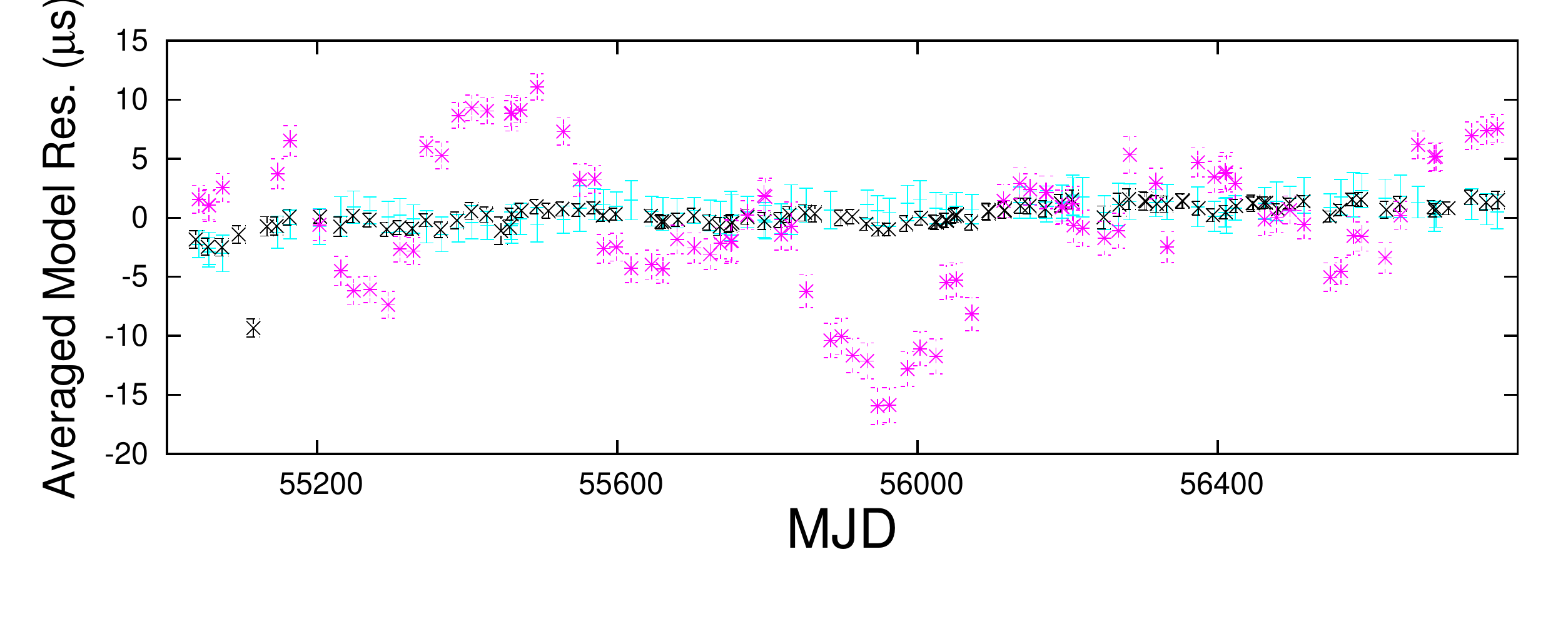}\\
\includegraphics[trim = 50 20 20 300, clip,width=150mm]{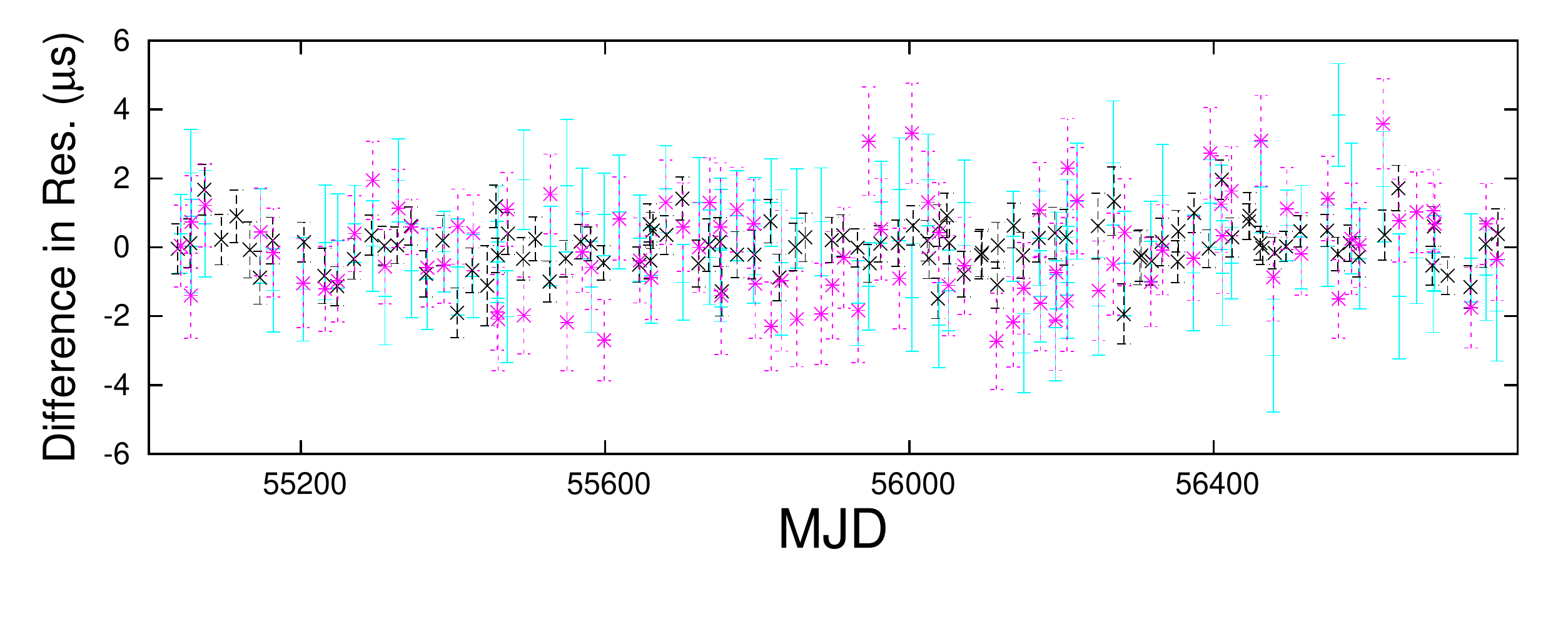}\\
\end{array}$
\end{center}
\caption{(Top Panel) Timing residuals after subtracting the maximum likelihood timing model, including DM variations obtained from the profile domain analysis performed in scenario~(S3).  Colours in this, and low panels, indicate the observing frequency of the data: 10\,cm (cyan points), 20\,cm (black points), and 40\,cm (magenta points).  (Middle panel)  Model residuals obtained by forming ToAs from the maximum-likelihood model profiles from scenario~(S3), and subtracting the same timing-model as in the top panel.  The time-variable broadening introduced by the scattering model results in completely consistent structure as the observed ToAs.  (Bottom panel) The difference between the residuals in the top- and middle-panels.  The `corrected` residuals are now consistent with white-noise, however, we stress this is only to be used figuratively.  The residuals obtained by subtracting the maximum-likelihood timing offsets introduced by scattering should not be used for analysis as they do not propagate the uncertainties in the scattering parameters through to the timing uncertainties. }
\label{figure:J1643ResPlot}
\end{figure*}

However, this implies we must also ask whether the model used for scattering is sufficient.  If the thin screen assumption is not valid, then bias will still enter into our parameter estimates.  In Fig.~\ref{figure:J1643ResPlot} we show the timing residuals after subtracting the timing model given in Table~\ref{Table:1643Params} obtained from our analysis of scenario~(S3).  In the middle panel we then show the timing residuals obtained from the maximum likelihood model profile data that includes variation in the pulse width due to scattering.  In both cases significant additional structure in the 40~cm data (magenta points) can clearly be seen as  a result of the scattering.  In the bottom panel of Fig.~\ref{figure:J1643ResPlot}  we then show the difference between the first two panels.  The residuals, after subtracting out delays that result from scattering variations, are consistent with white noise.  We stress that this process is performed purely for visualization of the profile domain model.  As discussed, simply subtracting the delay induced by scattering is not an appropriate means of correcting for scattering variations.

\section{Conclusions}
\label{Section:Conclusions}

Over the last several years different groups worldwide have sought to improve the precision with which pulsar timing can be performed by incorporating data across wide band widths, extending to low radio frequencies ($<$ 1~GHz).  The principal aim was to better correct for DM variations, which are known to dominate the noise budget for the majority of millisecond pulsars being used as part of an ongoing effort to detect gravitational waves in the nHz band.

As better models for these DM variations were developed, excess noise in this low frequency data has been observed in an increasing number of pulsars.  In \cite{2016MNRAS.458.2161L}, four pulsars in the International Pulsar Timing Array data release were found to have `band noise', including PSR  J1643$-$1224.  Even the most stable millisecond pulsar known to date, PSR J1909$-$3744, has been observed to suffer from band noise in the low frequency data, which, unless modeled appropriately, decreased the sensitivity of the data set to gravitational waves by a factor of two \citep{2015Sci...349.1522S}.

These band-noise models described the excess noise as time-correlated shifts in the arrival time of the pulses. Ultimately, however, if the origin of this noise is the result of time-variable broadening of the pulse profile due to scattering in the IISM, describing it in terms of shifts in the ToAs will be sub-optimal compared to appropriately modeling the changes in the pulse shape.

We have described a statistically robust approach to incorporating models for scattering into pulsar timing analysis  by extending the existing profile-domain timing framework. This approach makes it possible to simultaneously estimate temporal variations in both the DM and scattering, together with a model for the pulse profile that includes smooth evolution as a function of frequency, the pulsars timing model, and models for pulse jitter. 

We have shown that unless both the DM variations and scattering are sampled together in the timing analysis one can significantly underestimate the uncertainties in both the DM, and the arrival time of the pulse, leading to bias in the timing parameters.  This means that fixing the scattering timescales and simply `correcting' the ToAs, as has been suggested as a means of handling scattering variations in the literature,  is not a viable approach to performing a robust analysis of low frequency pulsar timing data.  We also showed, however, that the true probability density function (PDF) of the arrival time of the pulse can be highly non-Gaussian when scattering is only marginally detected.  In this case,  the typical procedure for forming a ToA, which assumes a Gaussian approximation to the true PDF, is inappropriate.  However, by performing a profile domain analysis the non-Gaussianity of the PDF is automaticlaly incorporated into the final parameter estimates.

We have applied our method using a new, publicly available, GPU accelerated code both to simulations, and observations of the millisecond pulsar PSR J1643$-$1224 that cover a frequency range from 700~MHz up to 3800~MHz over a period of 4.7~yr.  This particular MSP was chosen because it is known to display significant scattering variability compared to typical MSPs.  The techniques we have introduced will be applied to a broader, and more typical population of pulsars in subsequent work.  When including the low frequency data in the PSR J1643$-$1224 analysis the parameter estimates for both the DM variations and timing model are significantly biased by the variations in the pulse width due to scattering.  Further, simply fixing the scattering timescales to maximum likelihood values - equivalent to fixing the time offset when making ToAs - results in almost a factor two underestimation in the uncertainties of the DM variations compared to the simultaneous analysis.  

If low-frequency observations from telescopes such as LOFAR, or the Square Kilometre Array are to be incorporated into pulsar timing analysis then the bias that results from failing to appropriately model scattering variations will become extremely important.  The increased sensitivity afforded by these instruments will likely require more complex models than simple thin-screen approximations for scattering.  Any differences between the true pulse-broadening function and the models used will manifest as additional band-dependent noise, down-weighting the low-frequency data.  Given finite observing time, one must ask how much of that time to spend observing at low-frequencies.  If we are able to completely model the variations in pulse shape, then low-frequency observations will naturally be of great benefit.  The actual improvement, however, will vary from pulsar to pulsar depending on the strength and complexity of the scattering variations.

For pulsars that are bright at high frequencies ($\ga 3$GHz) the optimal approach may well be to spend all available observing time in a regime where IISM effects are negligible.  Such an approach has already resulted in the most sensitive limit on an isotropic gravitational wave background using a pulsar timing array to date \citep{2015Sci...349.1522S}. However, pulsars are known to have steep spectral indices ($S_{\nu} \propto \nu^{-1.8}$ on average, \citealt{2000A&AS..147..195M}),  and so in many cases observations at lower frequencies are necessary.  In such instances our ability to propagate our uncertainties in the scattering variability through to all other parameters will be critical, especially in the context of gravitational wave astronomy.  This is all handled automatically as part of a profile domain analysis, which will inevitably form the basis of modern, broad-band  pulsar timing analysis moving forward.

\section{Acknowledgements}

The Parkes radio telescope is part of the Australia Telescope National Facility which is funded by the Commonwealth of Australia for operation as a National Facility managed by Commonwealth Science and Industrial Research Organization (CSIRO).  We thank all of the observers, engineers, and Parkes observatory staff members who have assisted with the observations reported in this paper.

LL was supported by a Junior Research Fellowship at Trinity Hall College, Cambridge University.
SO was supported by the Alexander von Humboldt Foundation.

We thank Stephen Taylor for his assistance in building the python software used in this paper.

\bibliographystyle{mn2e}
\bibliography{references}

\begin{thebibliography}{37}
\expandafter\ifx\csname natexlab\endcsname\relax\def\natexlab#1{#1}\fi

\bibitem[{{Arzoumanian} {et~al}\mbox{.}(2015){Arzoumanian}, {Brazier},
  {Burke-Spolaor}, {Chamberlin}, {Chatterjee}, {Christy}, {Cordes}, {Cornish},
  {Crowter}, {Demorest}, {Dolch}, {Ellis}, {Ferdman}, {Fonseca},
  {Garver-Daniels}, {Gonzalez}, {Jenet}, {Jones}, {Jones}, {Kaspi}, {Koop},
  {Lam}, {Lazio}, {Levin}, {Lommen}, {Lorimer}, {Luo}, {Lynch}, {Madison},
  {McLaughlin}, {McWilliams}, {Nice}, {Palliyaguru}, {Pennucci}, {Ransom},
  {Siemens}, {Stairs}, {Stinebring}, {Stovall}, {Swiggum}, {Vallisneri}, {van
  Haasteren}, {Wang}, \& {Zhu}}]{2015ApJ...813...65T}
{Arzoumanian} Z. {et~al.}, 2015, \apj, 813, 65

\bibitem[{{Cognard} {et~al}\mbox{.}(2013){Cognard}, {Theureau}, {Guillemot},
  {Liu}, {Lassus}, \& {Desvignes}}]{2013sf2a.conf..327C}
{Cognard} I., {Theureau} G., {Guillemot} L., {Liu} K., {Lassus} A., {Desvignes}
  G., 2013, in SF2A-2013: Proceedings of the Annual meeting of the French
  Society of Astronomy and Astrophysics, {Cambresy} L., {Martins} F., {Nuss}
  E., {Palacios} A., eds., pp. 327--330

\bibitem[{{Coles} {et~al}\mbox{.}(2015){Coles}, {Kerr}, {Shannon}, {Hobbs},
  {Manchester}, {You}, {Bailes}, {Bhat}, {Burke-Spolaor}, {Dai}, {Keith},
  {Levin}, {Oslowski}, {Ravi}, {Reardon}, {Toomey}, {van Straten}, {Wang},
  {Wen}, \& {Zhu}}]{2015arXiv150607948C}
{Coles} W.~A. {et~al.}, 2015, ArXiv e-prints

\bibitem[{{Coles} {et~al}\mbox{.}(2010){Coles}, {Rickett}, {Gao}, {Hobbs}, \&
  {Verbiest}}]{2010ApJ...717.1206C}
{Coles} W.~A., {Rickett} B.~J., {Gao} J.~J., {Hobbs} G., {Verbiest} J.~P.~W.,
  2010, \apj, 717, 1206

\bibitem[{{Cordes}(1978)}]{1978ApJ...222.1006C}
{Cordes} J.~M., 1978, \apj, 222, 1006

\bibitem[{{Cordes}, {Shannon} \& {Stinebring}(2015){Cordes}, {Shannon}, \&
  {Stinebring}}]{2015arXiv150308491C}
{Cordes} J.~M., {Shannon} R.~M., {Stinebring} D.~R., 2015, ArXiv e-prints

\bibitem[{{Dai} {et~al}\mbox{.}(2015){Dai} {et~al.}}]{2015MNRAS.449.3223D}
{Dai} S., {et~al.}, 2015, \mnras, 449, 3223

\bibitem[{{Demorest} {et~al}\mbox{.}(2013){Demorest}
  {et~al.}}]{2013ApJ...762...94D}
{Demorest} P.~B., {et~al.}, 2013, The Astrophys. J., 762, 94

\bibitem[{{Feroz} \& {Hobson}(2008)}]{2008MNRAS.384..449F}
{Feroz} F., {Hobson} M.~P., 2008, Mon. Not. R. Astron. Soc., 384, 449

\bibitem[{{Feroz}, {Hobson} \& {Bridges}(2009){Feroz}, {Hobson}, \&
  {Bridges}}]{2009MNRAS.398.1601F}
{Feroz} F., {Hobson} M.~P., {Bridges} M., 2009, Mon. Not. R. Astron. Soc., 398,
  1601

\bibitem[{{Handley}, {Hobson} \& {Lasenby}(2015){Handley}, {Hobson}, \&
  {Lasenby}}]{2015arXiv150201856H}
{Handley} W.~J., {Hobson} M.~P., {Lasenby} A.~N., 2015, ArXiv e-prints

\bibitem[{{Hankins} \& {Cordes}(1981)}]{1981ApJ...249..241H}
{Hankins} T.~H., {Cordes} J.~M., 1981, \apj, 249, 241

\bibitem[{{Hewish} {et~al}\mbox{.}(1968){Hewish}, {Bell}, {Pilkington},
  {Scott}, \& {Collins}}]{1968Natur.217..709H}
{Hewish} A., {Bell} S.~J., {Pilkington} J.~D.~H., {Scott} P.~F., {Collins}
  R.~A., 1968, \nat, 217, 709

\bibitem[{{Keith} {et~al}\mbox{.}(2013){Keith} {et~al.}}]{2013MNRAS.429.2161K}
{Keith} M.~J., {et~al.}, 2013, Mon. Not. R. Astron. Soc., 429, 2161

\bibitem[{{Kramer} {et~al}\mbox{.}(2006){Kramer}, {Stairs}, {Manchester},
  {McLaughlin}, {Lyne}, {Ferdman}, {Burgay}, {Lorimer}, {Possenti}, {D'Amico},
  {Sarkissian}, {Hobbs}, {Reynolds}, {Freire}, \&
  {Camilo}}]{2006Sci...314...97K}
{Kramer} M. {et~al.}, 2006, Science, 314, 97

\bibitem[{{Lee} {et~al}\mbox{.}(2014){Lee}, {Bassa}, {Janssen}, {Karuppusamy},
  {Kramer}, {Liu}, {Perrodin}, {Smits}, {Stappers}, {van Haasteren}, \&
  {Lentati}}]{2014MNRAS.441.2831L}
{Lee} K.~J. {et~al.}, 2014, \mnras, 441, 2831

\bibitem[{{Lentati}, {Alexander} \& {Hobson}(2015){Lentati}, {Alexander}, \&
  {Hobson}}]{2015MNRAS.447.2159L}
{Lentati} L., {Alexander} P., {Hobson} M.~P., 2015, \mnras, 447, 2159

\bibitem[{{Lentati} {et~al}\mbox{.}(2013){Lentati}, {Alexander}, {Hobson},
  {Taylor}, {Gair}, {Balan}, \& {van Haasteren}}]{2013PhRvD..87j4021L}
{Lentati} L., {Alexander} P., {Hobson} M.~P., {Taylor} S., {Gair} J., {Balan}
  S.~T., {van Haasteren} R., 2013, Phys. Rev. D, 87, 104021

\bibitem[{{Lentati}, {Hobson} \& {Alexander}(2014){Lentati}, {Hobson}, \&
  {Alexander}}]{2014MNRAS.444.3863L}
{Lentati} L., {Hobson} M.~P., {Alexander} P., 2014, \mnras, 444, 3863

\bibitem[{{Lentati} \& {Shannon}(2015)}]{2015MNRAS.454.1058L}
{Lentati} L., {Shannon} R.~M., 2015, \mnras, 454, 1058

\bibitem[{{Lentati} {et~al}\mbox{.}(2016){Lentati}, {Shannon}, {Coles},
  {Verbiest}, {van Haasteren}, {Ellis}, {Caballero}, {Manchester},
  {Arzoumanian}, {Babak}, {Bassa}, {Bhat}, {Brem}, {Burgay}, {Burke-Spolaor},
  {Champion}, {Chatterjee}, {Cognard}, {Cordes}, {Dai}, {Demorest},
  {Desvignes}, {Dolch}, {Ferdman}, {Fonseca}, {Gair}, {Gonzalez}, {Graikou},
  {Guillemot}, {Hessels}, {Hobbs}, {Janssen}, {Jones}, {Karuppusamy}, {Keith},
  {Kerr}, {Kramer}, {Lam}, {Lasky}, {Lassus}, {Lazarus}, {Lazio}, {Lee},
  {Levin}, {Liu}, {Lynch}, {Madison}, {McKee}, {McLaughlin}, {McWilliams},
  {Mingarelli}, {Nice}, {Os{\l}owski}, {Pennucci}, {Perera}, {Perrodin},
  {Petiteau}, {Possenti}, {Ransom}, {Reardon}, {Rosado}, {Sanidas}, {Sesana},
  {Shaifullah}, {Siemens}, {Smits}, {Stairs}, {Stappers}, {Stinebring},
  {Stovall}, {Swiggum}, {Taylor}, {Theureau}, {Tiburzi}, {Toomey},
  {Vallisneri}, {van Straten}, {Vecchio}, {Wang}, {Wang}, {You}, {Zhu}, \&
  {Zhu}}]{2016MNRAS.458.2161L}
{Lentati} L. {et~al.}, 2016, \mnras, 458, 2161

\bibitem[{{Liu} {et~al}\mbox{.}(2014){Liu}, {Desvignes}, {Cognard}, {Stappers},
  {Verbiest}, {Lee}, {Champion}, {Kramer}, {Freire}, \&
  {Karuppusamy}}]{2014MNRAS.443.3752L}
{Liu} K. {et~al.}, 2014, \mnras, 443, 3752

\bibitem[{{Manchester}(2015)}]{2015IAUGA..2256190M}
{Manchester} R.~N., 2015, IAU General Assembly, 22, 2256190

\bibitem[{{Manchester} {et~al}\mbox{.}(2013){Manchester}, {Hobbs}, {Bailes},
  {Coles}, {van Straten}, {Keith}, {Shannon}, {Bhat}, {Brown}, {Burke-Spolaor},
  {Champion}, {Chaudhary}, {Edwards}, {Hampson}, {Hotan}, {Jameson}, {Jenet},
  {Kesteven}, {Khoo}, {Kocz}, {Maciesiak}, {Oslowski}, {Ravi}, {Reynolds},
  {Sarkissian}, {Verbiest}, {Wen}, {Wilson}, {Yardley}, {Yan}, \&
  {You}}]{2013PASA...30...17M}
{Manchester} R.~N. {et~al.}, 2013, \pasa, 30, 17

\bibitem[{{Narayan}(1992)}]{1992RSPTA.341..151N}
{Narayan} R., 1992, Philosophical Transactions of the Royal Society of London
  Series A, 341, 151

\bibitem[{Nelder \& Mead(1965)}]{NelderMead65}
Nelder J.~A., Mead R., 1965, Computer Journal, 7, 308

\bibitem[{{Os{\l}owski} {et~al}\mbox{.}(2011){Os{\l}owski}, {van Straten},
  {Hobbs}, {Bailes}, \& {Demorest}}]{2011MNRAS.418.1258O}
{Os{\l}owski} S., {van Straten} W., {Hobbs} G.~B., {Bailes} M., {Demorest} P.,
  2011, \mnras, 418, 1258

\bibitem[{{Pennucci}, {Demorest} \& {Ransom}(2014){Pennucci}, {Demorest}, \&
  {Ransom}}]{2014ApJ...790...93P}
{Pennucci} T.~T., {Demorest} P.~B., {Ransom} S.~M., 2014, \apj, 790, 93

\bibitem[{{Ransom} {et~al}\mbox{.}(2009){Ransom}, {Demorest}, {Ford},
  {McCullough}, {Ray}, {DuPlain}, \& {Brandt}}]{2009AAS...21460508R}
{Ransom} S.~M., {Demorest} P., {Ford} J., {McCullough} R., {Ray} J., {DuPlain}
  R., {Brandt} P., 2009, in American Astronomical Society Meeting Abstracts,
  Vol. 214, American Astronomical Society Meeting Abstracts 214, p. 605.08

\bibitem[{{Reardon} {et~al}\mbox{.}(2016){Reardon}, {Hobbs}, {Coles}, {Levin},
  {Keith}, {Bailes}, {Bhat}, {Burke-Spolaor}, {Dai}, {Kerr}, {Lasky},
  {Manchester}, {Os{\l}owski}, {Ravi}, {Shannon}, {van Straten}, {Toomey},
  {Wang}, {Wen}, {You}, \& {Zhu}}]{2016MNRAS.455.1751R}
{Reardon} D.~J. {et~al.}, 2016, \mnras, 455, 1751

\bibitem[{{Refregier}(2003)}]{2003MNRAS.338...35R}
{Refregier} A., 2003, \mnras, 338, 35

\bibitem[{{Shannon} {et~al}\mbox{.}(2015){Shannon}, {Ravi}, {Lentati}, {Lasky},
  {Hobbs}, {Kerr}, {Manchester}, {Coles}, {Levin}, {Bailes}, {Bhat},
  {Burke-Spolaor}, {Dai}, {Keith}, {Os{\l}owski}, {Reardon}, {van Straten},
  {Toomey}, {Wang}, {Wen}, {Wyithe}, \& {Zhu}}]{2015Sci...349.1522S}
{Shannon} R.~M. {et~al.}, 2015, Science, 349, 1522

\bibitem[{{Shannon} {et~al}\mbox{.}(2014){Shannon}
  {et~al.}}]{2014MNRAS.443.1463S}
{Shannon} R.~M., {et~al.}, 2014, Mon. Not. R. Astron. Soc., 443, 1463

\bibitem[{{Stappers} {et~al}\mbox{.}(2011){Stappers}, {Hessels}, {Alexov},
  {Anderson}, {Coenen}, {Hassall}, {Karastergiou}, {Kondratiev}, {Kramer}, {van
  Leeuwen}, {Mol}, {Noutsos}, {Romein}, {Weltevrede}, {Fender}, {Wijers},
  {B{\"a}hren}, {Bell}, {Broderick}, {Daw}, {Dhillon}, {Eisl{\"o}ffel},
  {Falcke}, {Griessmeier}, {Law}, {Markoff}, {Miller-Jones}, {Scheers},
  {Spreeuw}, {Swinbank}, {Ter Veen}, {Wise}, {Wucknitz}, {Zarka}, {Anderson},
  {Asgekar}, {Avruch}, {Beck}, {Bennema}, {Bentum}, {Best}, {Bregman},
  {Brentjens}, {van de Brink}, {Broekema}, {Brouw}, {Br{\"u}ggen}, {de Bruyn},
  {Butcher}, {Ciardi}, {Conway}, {Dettmar}, {van Duin}, {van Enst}, {Garrett},
  {Gerbers}, {Grit}, {Gunst}, {van Haarlem}, {Hamaker}, {Heald}, {Hoeft},
  {Holties}, {Horneffer}, {Koopmans}, {Kuper}, {Loose}, {Maat},
  {McKay-Bukowski}, {McKean}, {Miley}, {Morganti}, {Nijboer}, {Noordam},
  {Norden}, {Olofsson}, {Pandey-Pommier}, {Polatidis}, {Reich},
  {R{\"o}ttgering}, {Schoenmakers}, {Sluman}, {Smirnov}, {Steinmetz}, {Sterks},
  {Tagger}, {Tang}, {Vermeulen}, {Vermaas}, {Vogt}, {de Vos}, {Wijnholds},
  {Yatawatta}, \& {Zensus}}]{2011A&A...530A..80S}
{Stappers} B.~W. {et~al.}, 2011, Astronomy \& Astrophysics, 530, A80

\bibitem[{{Taylor}, {Ashdown} \& {Hobson}(2008){Taylor}, {Ashdown}, \&
  {Hobson}}]{2008MNRAS.389.1284T}
{Taylor} J.~F., {Ashdown} M.~A.~J., {Hobson} M.~P., 2008, \mnras, 389, 1284

\bibitem[{{Taylor}(1992)}]{1992RSPTA.341..117T}
{Taylor} J.~H., 1992, Royal Society of London Philosophical Transactions Series
  A, 341, 117

\bibitem[{{Zhu} {et~al}\mbox{.}(2015){Zhu}, {Stairs}, {Demorest}, {Nice},
  {Ellis}, {Ransom}, {Arzoumanian}, {Crowter}, {Dolch}, {Ferdman}, {Fonseca},
  {Gonzalez}, {Jones}, {Jones}, {Lam}, {Levin}, {McLaughlin}, {Pennucci},
  {Stovall}, \& {Swiggum}}]{2015ApJ...809...41Z}
{Zhu} W.~W. {et~al.}, 2015, \apj, 809, 41

\end{thebibliography}


\begin{thebibliography}{3}
\expandafter\ifx\csname natexlab\endcsname\relax\def\natexlab#1{#1}\fi

\bibitem[{{Manchester} {et~al}\mbox{.}(2013){Manchester}, {Hobbs}, {Bailes},
  {Coles}, {van Straten}, {Keith}, {Shannon}, {Bhat}, {Brown}, {Burke-Spolaor},
  {Champion}, {Chaudhary}, {Edwards}, {Hampson}, {Hotan}, {Jameson}, {Jenet},
  {Kesteven}, {Khoo}, {Kocz}, {Maciesiak}, {Oslowski}, {Ravi}, {Reynolds},
  {Sarkissian}, {Verbiest}, {Wen}, {Wilson}, {Yardley}, {Yan}, \&
  {You}}]{2013PASA...30...17M}
{Manchester} R.~N. {et~al.}, 2013, \pasa, 30, 17

\bibitem[{{Refregier}(2003)}]{2003MNRAS.338...35R}
{Refregier} A., 2003, \mnras, 338, 35

\bibitem[{{Shannon} {et~al}\mbox{.}(2014){Shannon}
  {et~al.}}]{2014MNRAS.443.1463S}
{Shannon} R.~M., {et~al.}, 2014, Mon. Not. R. Astron. Soc., 443, 1463

\end{thebibliography}


\begin{thebibliography}{5}
\expandafter\ifx\csname natexlab\endcsname\relax\def\natexlab#1{#1}\fi

\bibitem[{{Handley}, {Hobson} \& {Lasenby}(2015){Handley}, {Hobson}, \&
  {Lasenby}}]{2015arXiv150201856H}
{Handley} W.~J., {Hobson} M.~P., {Lasenby} A.~N., 2015, ArXiv e-prints

\bibitem[{{Kass} \& {Raftery}(1995)}]{bayesRef}
{Kass} R.~E., {Raftery} A.~E., 1995, Journal of the American Statistical
  Association, 90, 791

\bibitem[{{Neal}(2000)}]{2000physics...9028N}
{Neal} R.~M., 2000, ArXiv Physics e-prints

\bibitem[{{Refregier}(2003)}]{2003MNRAS.338...35R}
{Refregier} A., 2003, \mnras, 338, 35

\bibitem[{{Skilling}(2004)}]{2004AIPC..735..395S}
{Skilling} J., 2004, in American Institute of Physics Conference Series, Vol.
  735, American Institute of Physics Conference Series, {Fischer} R., {Preuss}
  R., {Toussaint} U.~V., eds., pp. 395--405

\end{thebibliography}

\end{document}